\documentclass[aps,prx,reprint,onecolumn,superscriptaddress]{revtex4-2}
\usepackage[T1]{fontenc}
\usepackage[utf8]{inputenc}
\usepackage[a4paper, margin=2cm]{geometry}
\usepackage{physics} 
\usepackage{amsthm} 
\usepackage{amssymb}
\usepackage{bbold} 
\usepackage{mathtools} 
\usepackage{graphicx} 
\usepackage[hidelinks]{hyperref}
\usepackage{subfig}
\usepackage{natbib}
\usepackage{slashed}
\usepackage{mathrsfs}
\usepackage{tikz}
\usetikzlibrary{calc,math,positioning}

\theoremstyle{plain}

\newcommand{\intp}{\hat{\mathcal{O}}}
\renewcommand{\norm}[2][]{#1\|#2#1\|}
\renewcommand{\abs}[2][]{#1|#2#1|}
\renewcommand{\ket}[2][]{#1|#2#1\rangle}

\newcommand{\braketmath}[3][]{#1\langle#2#1|#3#1\rangle}
\renewcommand{\braket}[4][]{#1\langle #2 #1| #3 #1| #4 #1\rangle}

\begin{document}

\title{Creation of Wave Packets for Quantum Chromodynamics on Quantum Computers}

\author{Matteo Turco}
\affiliation{Physics of Information and Quantum Technologies Group, Centro de Física e Engenharia de Materiais Avançados (CeFEMA), Lisbon, Portugal}
\affiliation{Laboratory of Physics for Materials and Emergent Technologies, Lisbon, Portugal}
\affiliation{Instituto Superior Técnico, Universidade de Lisboa, Lisbon, Portugal}
\affiliation{PQI -- Portuguese Quantum Institute, Lisbon, Portugal}

\author{Gonçalo Quinta}
\affiliation{Instituto de Telecomunicações, Lisbon, Portugal}

\author{João Seixas}
\affiliation{Physics of Information and Quantum Technologies Group, Centro de Física e Engenharia de Materiais Avançados (CeFEMA), Lisbon, Portugal}
\affiliation{Laboratory of Physics for Materials and Emergent Technologies, Lisbon, Portugal}
\affiliation{Departamento de Física, Instituto Superior Técnico, Universidade de Lisboa, Lisbon, Portugal}
\affiliation{PQI -- Portuguese Quantum Institute, Lisbon, Portugal}

\author{Yasser Omar}
\affiliation{Physics of Information and Quantum Technologies Group, Centro de Física e Engenharia de Materiais Avançados (CeFEMA), Lisbon, Portugal}
\affiliation{Laboratory of Physics for Materials and Emergent Technologies, Lisbon, Portugal}
\affiliation{Departamento de Matemática, Instituto Superior Técnico, Universidade de Lisboa, Lisbon, Portugal}
\affiliation{PQI -- Portuguese Quantum Institute, Lisbon, Portugal}

\begin{abstract}
One of the most ambitious goals of quantum simulation of
quantum field theory is the description of
scattering
in real time, which would allow not only for computation
of scattering amplitudes, but also for studying the collision
process step by step. The initial state of such simulation
is made of typically two wave packets of stable particles moving on top of the vacuum,
whose preparation is difficult.
Here we extend a previous work to
create wave packets of a general kind of particle from the vacuum
in lattice gauge theories in various dimensions,
including three-dimensional QCD for the first time.
The conceptual foundation of this approach is the
Haag-Ruelle scattering theory, and the only theoretical
limitation is given by the presence of massless particles. In the context
of digital quantum computation, the wave packet creation
from the vacuum is implemented
with the technique known as LCU (linear combination of unitaries).
The preparation is performed successfully upon measuring an
ancillary register with a certain
probability, which vanishes polynomially in the
lattice spacing, the wave-packet energy and the momentum narrowness.
\end{abstract}

\maketitle

\section{Introduction}
Quantum computation is a new promising paradigm to render
feasible some computational problems that are classically
too hard to be solved even with the huge amount of resources
of the best supercomputers. More precisely, quantum computers are
expected to have a truly strong impact in problems
that require an exponentially large amount of classical resources
and a polynomially large amount of quantum resources.
For these problems, having a proper quantum computer
would make the difference between solving relevant cases and
not solving them. It would not be just a matter of making the computation faster.
However,
instances of such problems are rare. Of the few known examples,
simulation of complex
quantum systems, especially of strongly correlated
many-body systems, is probably the most prominent.
Traditional approaches, numerical or analytical, fail except in a handful
of cases, or for limited classes of questions. For these reasons,
quantum and quantum-inspired approaches
are being widely studied~\cite{Buyens-Haegeman-VanAcoleyen-Verschelde-Verstraete_2013-12,
Rigobello-Notarnicola-Magnifico-Montangero_2105,
Marshall-Pooser-Siopsis-Weedbrook_1503,
Surace-Lerose_2011,
Karpov-Zhu-Heller-Heyl_2011,
Vovrosh-Mukherjee-Bastianello-Knolle_2204,
Belyansky-Whitsitt-Mueller-Fahimniya-Bennewitz-Davoudi-Gorshkov_2307,
Bennewitz-Ware-Schuckert-Lerose-Surace-Belyansky-Morong-Luo-De-Collins-Katz-Monroe-Davoudi-Gorshkov_2403,
Fauseweh-Zhu_2020-09,
Fauseweh_2024-12,
Daley-Bloch-Kokail-Flannigan-Pearson-Troyer-Zoller_2022-07,
Savage_2023-12,
Delgado-Granados-Krogmeier-Sager-Smith-Avdic-Hu-Sajjan-Abbasi-Smart-Narang-Kais-Schlimgen-Head-Marsden-Mazziotti_2024-06,
Toshio-Akahoshi-Fujisaki-Oshima-Sato-Fujii_2024-08}.

A quantum field theory is essentially a strongly correlated
quantum system with particular requirements on
the symmetries it enjoys, namely Poincaré invariance.
At a deep level, the correct implementation of
this symmetry group also seems to demand gauge invariance
to obtain nontrivial dynamics, at least in a three-dimensional space.
A particularly challenging case is
quantum chromodynamics (QCD), which
describes the strong interaction between quarks and gluons.
Differently from all the other sectors of the Standard Model of
particle physics,
a large part of its phenomenology is intrinsically nonperturbative,
whence analytical techniques can be used only to set up the
right framework for numerical calculations. The gauge group
is $SU(3)$, which increases computational costs
by several orders of magnitude with respect, for instance, to a $U(1)$ theory. Traditional lattice calculations
are well suited to address problems in the imaginary-time
formalism, such as prediction on particle masses and matrix elements,
and information on the real-time evolution can be extracted in some
cases~\cite{Luscher_1986}.
However, general nonperturbative studies of processes in
real-time evolution is one of the major open problems
of particle physics.

For the reasons just mentioned, considerable effort
within the quantum-computing and
the high-energy-physics communities aims at simulating real-time evolution of processes within
QCD~\cite{Bauer-etal_2204,
Avkhadiev-Shanahan-Young_2020,
Avkhadiev-Shanahan-Young_2209,
Klco-Savage-Stryker_1908,
Mueller-Tarasov-Venugopalan_1908,
Kreshchuk-Kirby-Goldstein-Beauchemin-Love_2002,
Echevarria-Egusquiza-Rico-Schnell_2011,
Qian-Basili-Pal-Luecke-Vary_2112,
Davoudi-Mueller-Powers_2208,
Atas-Zhang-Lewis-Jahanpour-Haase-Muschik_2102,
Atas-Haase-Zhang-Wei-Pfaendler-Lewis-Muschik_2207,
Kane-Grabowska-Nachman-Bauer_2212,
Lamm-Lawrence-Yamauchi_1908,
Mariani-Pradhan-Ercolessi_2301,
Gustafson-Zhu-Dreher-Linke-Meurice_2021,
Farrell-Chernyshev-Powell-Zemlevskiy-Illa-Savage_2207,
Farrell-Chernyshev-Powell-Zemlevskiy-Illa-Savage_2209,
Schuhmacher-Su-Osborne-Gandon-Halimeh-Tavernelli_2025-05}.
The resources required to perform a useful quantum calculation
in QCD are several orders of magnitude beyond the present
capabilities~\cite{Bauer-Nachman-Freytsis_2102},
but promising progress has been made in the last
years towards the realization of a large-scale, error-corrected
quantum computer, both on the experimental and the theoretical
side~\cite{Bluvstein-et-al_2023-12,
Google_2024-08}.
Although a full quantum computer is still out of reach,
it nonetheless makes
sense to analyze all the possibilities to obtain realistic
resource estimates, and determine which questions we will be
able to address. Furthermore, on the long way to address full QCD
there are many simpler theories, such as the Schwinger model,
that can be used to test
and benchmark new techniques, and to gain insight on some
aspects of QCD and quantum field theory in general.

Specifically, we consider scattering of
particles, a class of experiments which
is extremely important for our understanding
of fundamental physics. At very high energy,
QCD is perturbative. The Lehmann-Symanzik-Zimmermann (LSZ)
reduction formula~\cite{Lehmann-Symanzik-Zimmermann_1955} is a powerful tool
to study cross sections at asymptotic times in perturbation theory
and to obtain predictions to be tested against
experimental results. However, at low energies, perturbation theory
is no longer useful, so it is very challenging to
study, for example, pion-pion or proton-proton collisions
above the inelastic-scattering threshold.
The Haag-Ruelle
formalism~\cite{Haag_1958,
Ruelle_1962,
Duncan}
is another excellent framework in which to study
scattering. It was developed in the context of axiomatic
quantum field theory, and it provides the logical link
between the Wightman axioms and the LSZ
reduction formula. While being a conceptual milestone of
quantum field theory, it is not well suited for perturbative calculations,
this being the main reason why it was outshone by the
LSZ formalism. However, quantum computation may be
the missing element to exploit the Haag-Ruelle theory
in a practical way too, as we will shortly see.

Quantum simulation offers a possibility to
study the real-time evolution of scattering processes,
but compared to other applications,
the initial state preparation
is particularly difficult. In the framework of
the Hamiltonian formulation on a spatial lattice,
our goal is to prepare two wave packets of
stable particles on top of the vacuum.
If the goal is just to
compute scattering amplitudes, wave packet preparation can be avoided~\cite{Li-Lai-Wang-Xing_2301,
Kreshchuk-Vary-Love_2310}.
However, simulation of
wave-packet collisions would allow for much more than computation of
scattering amplitudes. We could directly monitor the dynamics
step by step, and gain insight on what happens during the collision,
which is not easily accessible even in perturbative regimes.

Concerning the preparation of wave packets,
the key step in the first proposed
approaches consists of a modified adiabatic transformation
to turn wave packets of the free theory into
wave packets of the interacting
theory~\cite{Jordan-Lee-Preskill_1111,
Jordan-Lee-Preskill_1112,
Jordan-Lee-Preskill_1404,
Brennen-Rohde-Sanders-Singh_1412,
Barata-Mueller-Tarasov-Venugopalan_2012}.
Since in the free theory there are no states corresponding
to composite particles, this method can only be used
for particles associated with elementary degrees of freedom.
As a consequence of confinement in QCD, no quarks or gluons
appear in the spectrum, and all the particles (hadrons) are composite.
Therefore, the approach used
in~\cite{Jordan-Lee-Preskill_1111,
Jordan-Lee-Preskill_1112,
Jordan-Lee-Preskill_1404,
Brennen-Rohde-Sanders-Singh_1412,
Barata-Mueller-Tarasov-Venugopalan_2012}
does not apply to QCD.

To circumvent the problem, other strategies have to be followed.
Several ideas have been discussed and explored numerically
in one dimension.
In~\cite{Liu-Li-Zheng-Yuan-Sun_2021-09,Zemlevskiy_2024-11},
variational algorithms are used to obtain wave packets
in a scalar-field theory.
Variational approaches are also used in the Schwinger model
in~\cite{Farrell-Illa-Ciavarella-Savage_2401}.
Preparation of $W$ states combined
with variational techniques is used in~\cite{Farrell-Zemlevskiy-Illa-Preskill_2025-05},
which works well in models with
a short correlation length and, thus, far from the continuum limit.
In the context of the Abelian gauge theories $Z_2$ and $U(1)$,
ideas based on
ansatz-optimization~\cite{Davoudi-Hsieh-Kadam_2024-02,
Davoudi-Hsieh-Kadam_2025-05}
or quantum subspace expansion~\cite{Chai-Guo-Kuhn_2025-05} have been proposed
to obtain operators that create
mesonic wave packets from the interacting vacuum,
similarly to what is done
in~\cite{Rigobello-Notarnicola-Magnifico-Montangero_2105}
using tensor networks and in~\cite{Chai-Crippa-Jansen-Kuhn-Pascuzzi-Tacchino-Tavernelli_2312}
for the Thirring model.
Generalization of these methods to more
complicated cases, such as mesonic and baryonic
particles in multidimensional non-Abelian theories,
is possible in principle, but it has to be showed that these methods
remain efficient when more complicated
structures are considered, and when larger wave packets in coordinate space
are needed to obtain more energetic wave packets with a narrower
distribution in momentum space in the continuum limit.
In~\cite{Farrell-Illa-Ciavarella-Savage_2401}, another
idea is proposed for theories displaying confinement:
wave packets are adiabatically transformed from
the strong-coupling limit to the full theory, rather than
from the free theory. This should avoid the issue
explained in the previous paragraph, but more investigation
is required. A possible problem
is that in the continuum limit a theory like QCD is weakly coupled
due to asymptotic freedom, and the adiabatic transformation
may be difficult to implement.

A general approach, based on the Haag-Ruelle scattering theory,
has been introduced in our previous work~\cite{Turco-Quinta-Seixas-Omar_2305}, where we described how to obtain creation
operators for elementary and composite particles in
interacting scalar-field theories.
These operators act
on the interacting vacuum, so efficient interacting-vacuum preparation
is assumed to be available. Another required ingredient is
knowledge of the spectrum.
The only theoretical limitation of this strategy is the existence
of an isolated mass shell, which should be guaranteed by a finite mass gap
in the continuum limit. Apart from this, it is easily adaptable to different
theories, different formulations, and different particles,
as we show in this work. The creation operators are
obtained by appropriately smearing interpolating operators
for a given particle over space and time.
Its asymptotic efficiency can be shown
in the continuum limit and for high-energy wave packets
with narrow distribution in momentum space. The main difficulty
is that it only succeeds upon measuring
an ancillary register of qubits with a certain outcome.
The success probability is polynomially vanishing in the
lattice spacing, the wave-packet energy and the momentum narrowness.

In the present work, we show how to apply the
method based on the Haag-Ruelle formalism~\cite{Turco-Quinta-Seixas-Omar_2305}
to QCD. We
provide the first description of an efficient strategy to
prepare wave packets in this theory, assuming there
is an efficient way to prepare the vacuum state.
Apart from addressing theories different from
our previous work~\cite{Turco-Quinta-Seixas-Omar_2305},
the new aspects we address here
concern how to deal with particles with spin one-half and higher,
how to write interpolating operators in terms of qubit
operators, and estimating the success probability for
different particles.
To achieve this, we bring together
tools and concepts from different areas of research developed over
several decades, ranging from axiomatic quantum field theory,
traditional lattice field theory and Hamiltonian formulation of lattice
theories to quantum computation.
As a warm up, we consider $U(1)$, $SU(2)$ and $SU(3)$ theories
in one dimension. We use them to introduce
some building blocks required for three-dimensional QCD.
Moreover, the simulation of these theories
requires much less resources than full QCD, and
they are often used as toy models to test ideas.
Treating them separately from QCD may be useful
for implementations in the near or mid-term future.
We focus, for simplicity, on the well-known pions and nucleons.
Other kinds of particles can be treated in a completely
analogous way.

It is not clear which formulation of lattice QCD will
be best suited for quantum simulation, but a common choice has been
the Kogut-Susskind formulation, a choice we henceforth follow.
Other possibilities are the loop-string-hadron
formulation~\cite{Raychowdhury-Stryker_2018-12,
Raychowdhury-Stryker_2019-12,
Kadam-Raychowdhury-Stryker_2022-12,
Mathew-Raychowdhury_2024-04},
or the Wilson-fermion
formulation~\cite{Zache-Hebenstreit-Jendrzejewski-Oberthaler-Berges-Hauke_2018-02}.
The former is based on the Kogut-Susskind formulation, so developing
tools for the Kogut-Susskind formulation is typically necessary
or useful also for the loop-string-hadron one. Porting
the tools developed in this paper to the Wilson-fermion formulation
is straightforward.
Another option is to use the orbifold
formulation~\cite{Bergner-Hanada-Rinaldi-Schafer_2024-01,
Halimeh-Hanada-Matsuura-Nori-Rinaldi-Schafer_2024-11},
which can also be connected to the Kogut-Susskind framework.
Therefore, the choice of the Kogut-Susskind formulation is actually quite general.

The rest of this paper is organized as follows. In
section~\ref{2}, we review the approach presented
in~\cite{Turco-Quinta-Seixas-Omar_2305}, and we extend
it to particles of general spin. In section~\ref{3}
we give the lattice operators that are necessary for the
creation of wave packets in one-dimensional lattice gauge theories
$U(1)$, $SU(2)$ and $SU(3)$. In section~\ref{4}
we show how to implement the same operators in
terms of qubit operators.
Finally, in section~\ref{5} we consider two-flavoured QCD
in three dimensions, and in section~\ref{6} we give our conclusions.
In the appendix we review the generalized
superfast encoding of fermions onto qubits, and we also
show how to implement odd operators.

\section{Creation of Wave Packets from the Vacuum}
\label{2}
In this section we briefly review the main features
of the method proposed in~\cite{Turco-Quinta-Seixas-Omar_2305},
based on the Haag-Ruelle scattering theory,
to create wave packets of particles in an interacting field theory
from the vacuum of the theory. In ~\cite{Turco-Quinta-Seixas-Omar_2305} scalar-field theories were
considered, while in this paper we take at first a more general view. Then, at the
end of the section, we focus on spin-one-half particles, and
give some results that are necessary to determine the complexity
of the method.
The Haag-Ruelle
formulation~\cite{Haag_1958,Ruelle_1962} was originally developed in axiomatic
quantum field theory and recently formalized
for lattice systems~\cite{Bachmann-Dybalski-Naaijkens_1412}.
For a thorough review of the formulation in the continuum
see~\cite{Duncan}, whose extensive explanation goes
perhaps beyond the purposes of quantum simulation.
We invite the reader interested in quantum simulation
to go through~\cite{Turco-Quinta-Seixas-Omar_2305},
which also includes a more concise review of the Haag-Ruelle formalism.

For our purposes, the relevant result is that
we can create wave packets of
a certain particle using operators of the form
\begin{equation}
\hat{a}_{\psi}^{\dagger}=\int\!d^Dx\,\psi(x)\hat{\mathcal{O}}(x),
\label{continuum-HR-creation}
\end{equation}
where $D=d+1$ is the number of spacetime dimensions,
$x$ is a spacetime coordinate,
$\psi$ is a Schwartz function, and $\hat{\mathcal{O}}$
is an interpolating Heisenberg operator for the particle we want
to create.
The function $\psi$ and the operator
$\hat{\mathcal{O}}$ have to be properly chosen
considering the spectrum and the symmetries of the theory.
The operator in equation~\eqref{continuum-HR-creation} is sometime
called a Haag-Ruelle creation operator.
For proper choices of $\psi$ and $\hat{\mathcal{O}}$,
the state $\hat{a}_{\psi}^{\dagger}\ket{\Omega}$,
where $\ket{\Omega}$ denotes the vacuum, is a one-particle state.
To create multiple wave packets in different space
regions, we just need to apply multiple
Haag-Ruelle creation operators with spatially
separated functions $\psi_1,\,\psi_2,\dots$
Intuitively, the function $\psi$ is responsible for selecting one-particle
states and cutting off multiparticle states.
The operator $\hat{\mathcal{O}}$ is responsible
for selecting the desired kind of particle. Some examples
should help to understand these points.

\begin{figure}
\centering
\subfloat[][\label{joint-spectrum}A theory with only one
particle of mass $m$.]{\includegraphics{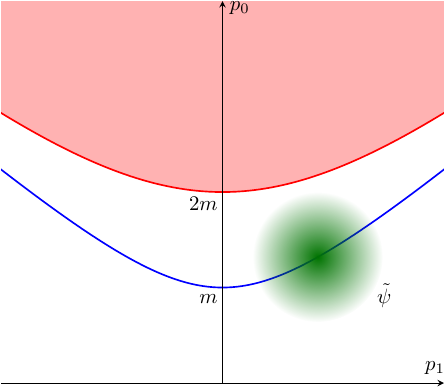}} \qquad
\subfloat[][\label{QCD-spectrum}Two-flavoured QCD.]
{\includegraphics{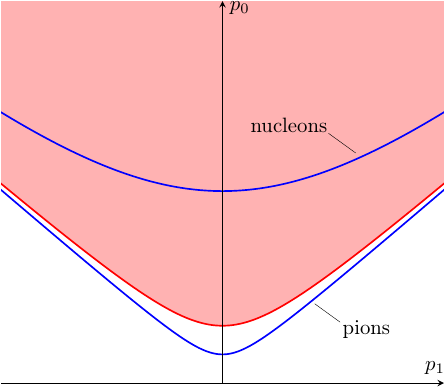}}
\caption{Two examples of the joint energy-momentum spectrum in continuum
quantum field theory. The blue lines represent
the one-particle mass hyperboloids, the red regions
are the multiparticle continua, and the green
spot represents the support of the function $\tilde{\psi}$.
For visual clarity, only one spatial component is shown.}
\end{figure}

To begin, suppose we have a theory
with only one particle of mass $m$ in the spectrum.
The momentum operators $P_0=H$ and $P_i$, $i=1,\dots,d$,
commute with each other, so we can consider their joint spectrum.
This is made of three separate sectors:
\begin{itemize}
\item the vacuum, corresponding to the origin, $p_{\mu}=0$;
\item the one-particle mass hyperboloid, given
by the points such that $p_{\mu}p^{\mu}=m^2$;
\item the multiparticle continuum, defined by
$p_{\mu}p^{\mu}\ge4m^2$.
\end{itemize}
The function $\psi$ is chosen as the Fourier
transform of a function $\tilde{\psi}$ with support
overlapping only with the one-particle mass hyperboloid,
and not with the multiparticle continuum, as
depicted in figure~\ref{joint-spectrum}.

The interpolating operator $\hat{\mathcal{O}}$ should
carry the same quantum numbers as the particle.
More precisely, if $\ket{\alpha}$ is a one-particle
state and $\ket{\Omega}$ the vacuum, then
the condition for $\hat{\mathcal{O}}$ to be an interpolating
operator is
\begin{equation}
\braket{\alpha}{\hat{\mathcal{O}}}{\Omega}\ne0.
\end{equation}

As a second example, suppose there are two particles of masses
$m_1$ and $m_2$. Our goal is to create wave packets of
a definite kind of particle, and not a mixture of the two.
If $m_2$ lies halfway between $m_1$ and $2m_1$, this can be achieved
by choosing $\tilde{\psi}$ with support only on the mass hyperboloid
of one kind of particle, namely either around $p_{\mu}p^{\mu}=m_1^2$
or $p_{\mu}p^{\mu}=m_2^2$. If $m_2$ is close to $m_1$,
close to $2m_1$, or is
larger than $2m_1$, then the operator $\hat{\mathcal{O}}$ must be
chosen to be an interpolating operator only for one kind of particle,
and not for the other. For example, we could have
\begin{equation}
\braket{\alpha_1}{\hat{\mathcal{O}}}{\Omega}\ne0,\qquad
\braket{\alpha_2}{\hat{\mathcal{O}}}{\Omega}=0,
\end{equation}
where $\ket{\alpha_i}$ is a state of particle $i$.
The way to identify particles, and hence appropriate interpolating
operators, is through symmetry considerations: the interpolating
operator should carry the same quantum numbers as the particle in consideration.

Finally, we can consider QCD with two flavours of
quarks. The low-lying spectrum is depicted in
figure~\ref{QCD-spectrum}, where only one spatial
component is shown for visual clarity.
The fact that
the one-nucleon mass hyperboloid lies in the multipion
continuum is not a problem because the two kind of particles
have different spins and isospins.

On a lattice, the same ideas apply with due modifications.
Coordinate space and momentum space are discretized, while time
is left continuous. Spectra are distorted, but should still contain
separate one-particle and multiparticle sectors.
We may then rewrite equation~\eqref{continuum-HR-creation}
on a lattice with spacing $a$ as
\begin{equation}
\hat{a}_{\psi}^{\dagger}=\sum_{\mathbf{x}}a^d\int_{-\infty}^{+\infty}\!dt\,
\psi(t,\mathbf{x})e^{iHt}\hat{\mathcal{O}}(\mathbf{x})e^{-iHt},
\label{lattice-HR-creation}
\end{equation}
where $\hat{\mathcal{O}}(\mathbf{x})$ is the Schrödinger representation
of $\hat{\mathcal{O}}(x)$, and $x=(t,\mathbf{x})$.
When moving from the continuum to the lattice,
many symmetries are lost, and new particles may appear.
As a consequence, there arises the problem of
identifying particles and using the correct lattice interpolators.
We deal with this more in detail in section~\ref{5}.

The lattice theory has to be mapped onto a qubit system,
with all due truncation and
digitization~\cite{Klco-Savage_1808,Tong-Albert-McClean-Preskill-Su_2110}.
Also the time integral in equation~\eqref{lattice-HR-creation}
has to be approximated by a discrete and truncated summation
in steps of $\delta_t$ over $N$ points,
\begin{equation}
\hat{a}_{\psi}^{\dagger}=\sum_{\mathbf{x}}\sum_{i=1}^Na^d\delta_t
\psi(t_i,\mathbf{x})e^{iHt_i}\hat{\mathcal{O}}(\mathbf{x})e^{-iHt_i}.
\end{equation}
The required number of time samples can be shown to grow like
\begin{equation}
N=O\bigg(
\frac{T}{\sqrt{\epsilon}}\sqrt{\norm[\big]{\big[H,[H,\intp(\mathbf{x})]\big]}}
\bigg),
\end{equation}
where $T$ is the size of $\psi$ in the time direction,
and $\epsilon$ is the error introduced by the time-integral
discretization. On a quantum computer, we can only
implement operators with $\norm{\intp}=O(1)$. Assuming
the Hamiltonian is a sum of local terms,
$H=\sum_{\mathbf{x}} H(\mathbf{x})$, we conclude that
\begin{equation}
N=O\bigg(
\frac{T}{\sqrt{\epsilon}}\norm{H(\mathbf{x})}
\bigg).
\end{equation}
Notice that $N$ does not depend on the lattice size.

Once we know how to implement $\intp$
on a quantum computer, the creation operator
$\hat{a}_{\psi}^{\dagger}$ can be implemented by means
of a linear combination of unitaries
(LCU)~\cite{Childs-Kothari-Somma_1511}.
If $S$ is the number of lattice sites in the support
of $\psi$, this requires an ancillary register of
$\log_2(NS)$ qubits, $NS$ calls to
controlled versions of $\intp$, and time evolution
for a time $O(T)$. The time evolution is likely the most
expensive part in the implementation of $\hat{a}_{\psi}^{\dagger}$,
since $N$ depends only on the local terms of the Hamiltonian,
while the time evolution depends on the whole Hamiltonian.
Also, the local operators $\intp$ we consider in this
work are not hard to implement.

The interpolator $\intp(\mathbf{x})$ is typically not unitary,
so it has to be block encoded. Since any operator
on a qubit system can be
expanded as a linear combination of unitary operators,
in general we can use LCU,
\begin{equation}
\intp(\mathbf{x})=\sum_{j=1}^Lc_j\mathcal{U}_j(\mathbf{x}),\qquad
\sum_j\abs{c_j}=C,\qquad
c_j\in\mathbb{C}.
\label{LCU}
\end{equation}
In the cases we consider below, $L$ is a constant depending
only on the kind of particle and the theory. Another
$\log_2L$ ancillary qubits are required to implement
$\intp(\mathbf{x})$. The quantity $C$ depends on $L$ and on the
mass dimension of the interpolator.

Overall, $\hat{a}_{\psi}^{\dagger}$ is successfully implemented
upon measurement of the ancillary register with probability
\begin{equation}
\rho=\bigg(
\frac{\norm[\big]{\hat{a}_{\psi}^{\dagger}\ket{\Omega}}}{\alpha}
\bigg)^2,\qquad\text{where }
\alpha=C\sum_{\mathbf{x},t}a^d\delta_t\abs{\psi(t,\mathbf{x})}.
\label{success-probability}
\end{equation}
The success probability determines the efficiency
of our method for wave-packet preparation. If we want to create
two wave packets, with success probabilities $\rho_1$ and
$\rho_2$, we need to repeat the vacuum preparation and
wave-packet creation about $1/(\rho_1\rho_2)$ times
before succeeding. This can be quadratically improved
applying amplitude amplification~\cite{Brassard-Hoyer-Mosca-Tapp_00}.

If the operator $\intp(x)$ is a scalar field,
the analysis in~\cite{Turco-Quinta-Seixas-Omar_2305}
shows that the behavior of $\rho$ is
\begin{equation}
\rho\sim\frac{Z}{C^2}\frac{\delta_p^d}{\bar{E}},
\label{asymptotic-success-probability}
\end{equation}
where $\delta_p$ is the linear size of the support of $\tilde{\psi}$
in the spatial components of the momentum,
$\bar{E}$ is the energy of the particle,
and $Z$ is the squared renormalization constant
connecting $\intp(\mathbf{x})$ with its continuum normalization,
typically logarithmic in the lattice
spacing~\cite{Capitani_2002-11}.

In general, $\intp(x)$ can be a spinorial field (or another
kind of field), as is the case for the proton, and the result
has to be adapted. Assume that under a Lorentz transformation
in the continuum
we have $\intp_{\alpha}\to\sum_{\beta}S(\Lambda)_{\alpha\beta}
\intp_{\beta}$, and that the operators $\intp_{\alpha}$
interpolate for a particle of spin $J$ and mass $m$. There are
$2J+1$ one-particle states with momentum $\mathbf{k}$,
$\ket{\mathbf{k},j}$, with $j=-J,-J+1,\dots,J$.
Then we have
\begin{equation}
\norm[\big]{\hat{a}_{\alpha,\psi}^{\dagger}\ket{\Omega}}^2=
\int d^3k\abs[\big]{\tilde{\psi}(\mathbf{k})}^2\frac{m}{E(\mathbf{k})}
\sum_j
\abs[\Big]{\sum_{\beta}S(\Lambda_{\mathbf{k}})_{\alpha\beta}\sqrt{Z_{\beta}^j}}^2,
\label{general-norm}
\end{equation}
where $E(\mathbf{k})=\sqrt{\abs{\mathbf{k}}^2+m^2}$,
$\tilde{\psi}(\mathbf{k})=\tilde{\psi}(E(\mathbf{k}),\mathbf{k})$,
$\sqrt{Z_{\beta}^j}=\braket{\mathbf{0},j}{\intp_{\beta}(0)}{\Omega}$,
and $\Lambda_{\mathbf{k}}$ is the boost that sends $\mathbf{k}$
to $\mathbf{0}$. If $\tilde{\psi}$ is narrowly
concentrated around $\bar{\mathbf{k}}$, then we roughly
have $S(\Lambda_{\mathbf{k}})\sim S(\Lambda_{\bar{\mathbf{k}}})$
in equation~\eqref{general-norm}. Therefore,
in equation~\eqref{asymptotic-success-probability},
we can replace $Z$ with
\begin{equation}
Z\rightarrow
\sum_j
\abs[\Big]{\sum_{\beta}S(\Lambda_{\bar{\mathbf{k}}})_{\alpha\beta}\sqrt{Z_{\beta}^j}}^2.
\end{equation}

We consider now the example of a spinorial field
in the Dirac representation interpolating
for a spin-one-half particle with momentum $\bar{\mathbf{k}}=(k_x,0,0)$.
We have
\begin{equation}
S(\Lambda_{\bar{\mathbf{k}}})=
\begin{pmatrix}
\sqrt{\frac{\gamma+1}{2}}\mathbb{1}&\sqrt{\frac{\gamma-1}{2}}\sigma_x\\
\sqrt{\frac{\gamma-1}{2}}\sigma_x&\sqrt{\frac{\gamma+1}{2}}\mathbb{1}
\end{pmatrix},
\qquad
\gamma=\frac{E(\mathbf{\bar{\mathbf{k}}})}{m}.
\end{equation}
Then, for $k_x\gg m$, we have
\begin{equation}
\sum_{\beta}S(\Lambda_{\bar{\mathbf{k}}})_{\alpha\beta}\sqrt{Z_{\beta}^j}
=
\frac{1}{\sqrt{2}}\sqrt{\frac{k_x}{m}}\Big(
\sqrt{Z_1^j}-\sqrt{Z_4^j}
\Big)+
\frac{1}{2\sqrt{2}}\sqrt{\frac{m}{k_x}}\Big(
\sqrt{Z_1^j}+\sqrt{Z_4^j}
\Big)+\dots
\end{equation}
Notice that,
for a spinorial interpolator, and if $Z_1^j\ne Z_4^j$,
the term $k_x/m$ from this expansion compensates the term $\bar{E}$
in the numerator of
equation~\eqref{asymptotic-success-probability} for highly
energetic spin-one-half particles.
If instead $Z_1^j=Z_4^j$, then we need to consider the second
term in the expansion, which brings another suppression
factor of $1/\bar{E}$ in the success probability.

In this section we have discussed the general features
of how to prepare wave packets on quantum computers
using the Haag-Ruelle scattering theory, focusing on particles of
spin zero and one half. To continue the discussion, we need
to consider specific examples. In particular, we need to have
explicit expressions for the interpolating operators to
estimate the quantity $C$ in~\eqref{success-probability}.
This is the content of the following sections.

\section{Interpolating Operators in One-dimensional Lattice Gauge Theories}
\label{3}
In this and the following sections, we aim at showing
how to decompose interpolating operators into a
linear combination of unitaries on qubit systems,
providing some rough estimates on the resources needed and
the success probability.
We start by considering the simpler case of gauge
theories in one space dimension, which is enough to show all the
building blocks that are necessary in the physical case
of quantum chromodynamics.

The Kogut-Susskind Hamiltonians for lattice gauge theories $U(1)$,
$SU(2)$ and $SU(3)$ in one space dimension can be compactly written
as~\cite{Kan-Nam_2107}
\begin{equation}
	H=\sum_{x=0}^{N-1}\Big[\frac{i}{2a}\big(\xi^{\dagger}(x+1)
	U(x)\xi(x)-\xi^{\dagger}(x)U^{\dagger}(x)\xi(x+1)\big)+
	(-1)^xm_0\xi^{\dagger}(x)\xi(x)+\frac{ag_0^2}{2}E(x)^2
	\Big].
\end{equation}
When referring to the $U(1)$ theory, $\xi(x)$, $U(x)$ and $E(x)$ are one-component
fields. When referring to the $SU(2)$ theory, the same symbols mean
\begin{equation}
\xi(x)=
\begin{pmatrix}
	\xi_1(x)\\
	\xi_2(x)
\end{pmatrix},\qquad
U(x)=
\begin{pmatrix}
	U_{11}(x)	& U_{12}(x)\\
	U_{21}(x)	& U_{22}(x)
\end{pmatrix},\qquad
E(x)^2=\sum_{a=1}^3E_{\text{L}}^a(x)^2
=\sum_{a=1}^3E_{\text{R}}^a(x)^2.
\end{equation}
When referring to the $SU(3)$ theory, the symbols mean
\begin{equation}
\xi(x)=
\begin{pmatrix}
	\xi_1(x)\\
	\xi_2(x)\\
	\xi_3(x)
\end{pmatrix},\qquad
U(x)=
\begin{pmatrix}
	U_{11}(x)	& U_{12}(x)	& U_{13}(x)\\
	U_{21}(x)	& U_{22}(x) & U_{23}(x)\\
	U_{31}(x)	& U_{32}(x) & U_{33}(x)\\
\end{pmatrix},\qquad
E(x)^2=\sum_{a=1}^8E_{\text{L}}^a(x)^2
=\sum_{a=1}^8E_{\text{R}}^a(x)^2.
\end{equation}

In these formulations we use staggered fermions.
A relevant caveat is that only even lattice
translations can be interpreted as space translations
in the continuum. Translations that involve an odd number of lattice units
in one or more directions do not correspond to space translations
in the continuum, but to other transformations that are not necessarily
symmetries of the theory. Hence, the summation
over $\mathbf{x}$ in equation~\eqref{lattice-HR-creation} should
be in steps of two, as in
\begin{equation}
\hat{a}_{\psi}^{\dagger}=\sum_{x=0}^{\frac{N}{2}-1}2a\int_{-\infty}^{+\infty}\!dt\,
\psi(t,2x)e^{iHt}\hat{\mathcal{O}}(2x)e^{-iHt}.
\end{equation}

\subsection{$U(1)$ gauge theory}
For the $U(1)$ gauge theory, also known as the Schwinger model,
the fields satisfy the following anticommutation and commutation
relations:
\begin{equation}
\{\xi(x),\xi^{\dagger}(y)\}=\delta_{xy},\qquad
[U(x),E(y)]=\delta_{xy}U(x).
\end{equation}
The local Hilbert space of the gauge field can be expressed
in the electric-field basis $\{\ket{\mathcal{E}}:\mathcal{E}\in\mathbb{Z}\}$,
in which we have
\begin{equation}
E\ket{\mathcal{E}}=\mathcal{E}\ket{\mathcal{E}},\qquad
U\ket{\mathcal{E}}=\ket{\mathcal{E}-1}.
\end{equation}
In this model there are three stable mesons in the continuum limit,
with associated interpolating
operators~\cite{Banks-Susskind-Kogut-1976}
\begin{equation}
\bar{\psi}\psi,\qquad\psi^{\dagger}\gamma_5\psi,\qquad
i\bar{\psi}\gamma_5\psi.
\label{Schwinger-interpolators}
\end{equation}
The Dirac spinor $\psi$ for the quark field has two components
in one dimension.
In the staggered-fermion formulation, the upper component
is identified with fermionic modes at even sites, while
the lower component is identified with fermionic modes at odd sites.
Furthermore, the dimensionless field $\xi$ has to be
multiplied by a factor $1/\sqrt{a}$ to match the continuum
normalization, $\psi\sim\xi/\sqrt{a}$.
The operators in equation~\eqref{Schwinger-interpolators}
translate respectively to
\begin{align}
\intp_1(x)&=
\frac{1}{a}\Big[\xi^{\dagger}(x)\xi(x)-\xi^{\dagger}(x+1)\xi(x+1)\Big],\\
\intp_2(x)&=
\frac{1}{a}\Big[\xi^{\dagger}(x)U^{\dagger}(x)\xi(x+1)+\xi^{\dagger}(x+1)U(x)\xi(x)\Big],\\
\intp_3(x)&=
\frac{1}{a}\Big[\xi^{\dagger}(x)U^{\dagger}(x)\xi(x+1)-\xi^{\dagger}(x+1)U(x)\xi(x)\Big],
\end{align}
where we have dropped the factor $i$ in $\intp_3$ as it is
irrelevant for our purposes.

\subsection{$SU(2)$ gauge theory}
The anticommutation and commutation relations for the $SU(2)$ gauge theory
are
\begin{equation}
\{\xi_{\alpha}(x),\xi_{\beta}^{\dagger}(y)\}=\delta_{xy}\delta_{\alpha\beta},\qquad
[U_{\alpha\beta}(x),E_{\text{L}}^a(y)]=\delta_{xy}T_{\alpha\gamma}^aU_{\gamma\beta}(x),\qquad
[U_{\alpha\beta}(x),E_{\text{R}}^a(y)]=-\delta_{xy}U_{\alpha\gamma}(x)T_{\gamma\beta}^a,
\end{equation}
where $T^a=\sigma^a/2$. We adopt the Einstein summation convention for repeated
indices. Latin indices like
$a,b,c,\dots$ run over $\{1,2,3\}$ for $SU(2)$ and over $\{1,2,\dots,8\}$
for $SU(3)$, while Greek indices like $\alpha,\beta,\gamma,\dots$ run over
$\{1,2\}$ for $SU(2)$ and over $\{1,2,3\}$ for $SU(3)$.
In addition we have $U_{11}^{\dagger}(x)=U_{22}(x)$ and
$U_{21}^{\dagger}(x)=-U_{12}(x)$.

A basis for the local Hilbert space of the gauge field is given by
\begin{equation}
\ket{j,m_{\text{L}},m_{\text{R}}},\qquad
j=0,\frac{1}{2},1,\dots,\qquad
m_{\text{L}},m_{\text{R}}=-j,-j+1,\dots,j,
\end{equation}
in which the operators $E^2$, $E^3_{\text{L}}$ and $E^3_{\text{R}}$
are diagonal and give
\begin{gather}
E^2\ket{j,m_{\text{L}},m_{\text{R}}}=j(j+1)\ket{j,m_{\text{L}},m_{\text{R}}},\\
E_{\text{L}}^3\ket{j,m_{\text{L}},m_{\text{R}}}=m_{\text{L}}\ket{j,m_{\text{L}},m_{\text{R}}},
\qquad
E_{\text{R}}^3\ket{j,m_{\text{L}},m_{\text{R}}}=m_{\text{R}}\ket{j,m_{\text{L}},m_{\text{R}}}.
\end{gather}
The operators $U_{\alpha\beta}$ can be expressed as
\begin{multline}
U_{\alpha\beta}\ket{j,m_{\text{L}},m_{\text{R}}}=\\
c_{\alpha\beta}^-(j,m_{\text{L}},m_{\text{R}})
\ket{j-\frac{1}{2},m_{\text{L}}+\frac{3}{2}-\alpha,m_{\text{R}}+\frac{3}{2}-\beta}+
c_{\alpha\beta}^+(j,m_{\text{L}},m_{\text{R}})
\ket{j+\frac{1}{2},m_{\text{L}}+\frac{3}{2}-\alpha,m_{\text{R}}+\frac{3}{2}-\beta},
\end{multline}
where $c_{\alpha\beta}^{\pm}(j,m_{\text{L}},m_{\text{R}})$
are products of Clebsch-Gordan coefficients,
\begin{equation}
c_{\alpha\beta}^{\pm}(j,m_{\text{L}},m_{\text{R}})=
\sqrt{\frac{2j+1}{2j\pm1+1}}
\braketmath{j\pm\frac{1}{2},m_{\text{L}}+\frac{3}{2}-\alpha}{j,m_{\text{L}};\frac{1}{2},\frac{3}{2}-\alpha}
\braketmath{j\pm\frac{1}{2},m_{\text{R}}+\frac{3}{2}-\beta}{j,m_{\text{R}};\frac{1}{2},\frac{3}{2}-\beta}.
\end{equation}
We also have the decomposition
\begin{equation}
U_{\alpha\beta}=M_{\alpha}^{\text{L}}M_{\beta}^{\text{R}}\big(
J^-c_{\alpha\beta}^-+J^+c_{\alpha\beta}^+\big),
\qquad\text{(no summation over $\alpha$ and $\beta$)}
\label{transporter-SU2}
\end{equation}
with
\begin{align}
M_{\alpha}^{\text{L}}\ket{j,m_{\text{L}},m_{\text{R}}}&=
\ket{j,m_{\text{L}}+\frac{3}{2}-\alpha,m_{\text{R}}},\\
M_{\beta}^{\text{R}}\ket{j,m_{\text{L}},m_{\text{R}}}&=
\ket{j,m_{\text{L}},m_{\text{R}}+\frac{3}{2}-\beta},\\
J^{\pm}\ket{j,m_{\text{L}},m_{\text{R}}}&=
\ket{j\pm\frac{1}{2},m_{\text{L}},m_{\text{R}}},\\
c_{\alpha\beta}^{\pm}\ket{j,m_{\text{L}},m_{\text{R}}}&=
c_{\alpha\beta}^{\pm}(j,m_{\text{L}},m_{\text{R}})\ket{j,m_{\text{L}},m_{\text{R}}}.
\end{align}

Analogously to the $U(1)$ theory, the $SU(2)$ theory has
mesons~\cite{Hamer-1977}
\begin{align}
\intp_1(x)&=
\frac{1}{a}\Big[\xi^{\dagger}(x)\xi(x)-\xi^{\dagger}(x+1)\xi(x+1)\Big],
\label{meson1-SU2}\\
\intp_2(x)&=
\frac{1}{a}\Big[\xi^{\dagger}(x)U^{\dagger}(x)\xi(x+1)+\xi^{\dagger}(x+1)U(x)\xi(x)\Big],\\
\intp_3(x)&=
\frac{1}{a}\Big[\xi^{\dagger}(x)U^{\dagger}(x)\xi(x+1)-\xi^{\dagger}(x+1)U(x)\xi(x)\Big].
\label{meson3-SU2}
\end{align}
In addition, it has a baryon,
\begin{equation}
\intp_4(x)=
\frac{1}{a}\epsilon_{\alpha\beta}\xi_{\alpha}^{\dagger}(x)\xi_{\beta}^{\dagger}(x),
\end{equation}
where $\epsilon_{\alpha\beta}$ is the Levi-Civita symbol. However, the baryon is
expected to coincide with a meson in the continuum limit.

\subsection{$SU(3)$ gauge theory}
The $SU(3)$ case is a more complicated version of the $SU(2)$ one.
For the details see~\cite{Kan-Nam_2107}.
The basis states are labeled by eight quantum numbers
\begin{gather}
\ket{p,q,T_{\text{L}},T_{\text{L}}^z,Y_{\text{L}},T_{\text{R}},T_{\text{R}}^z,Y_{\text{R}}},\\
p,q=0,1,2,\dots,\qquad
T_{{\text{L}}/{\text{R}}}=0,\frac{1}{2},\dots,\frac{1}{2}(p+q),\\
T_{{\text{L}}/{\text{R}}}^z=
-\frac{1}{2}(p+q),-\frac{1}{2}(p+q+1),\dots,\frac{1}{2}(p+q),\qquad
Y_{{\text{L}}/{\text{R}}}=-\frac{1}{3}(q+2p),-\frac{1}{3}(q+2p+1),\dots,\frac{1}{3}(p+2q).
\end{gather}
Similarly to the $SU(2)$ case, in this basis we have the decomposition
\begin{equation}
U_{\alpha\beta}=M_{\alpha}^{\text{L}}M_{\beta}^{\text{R}}\big(
P^+\mathcal{C}_{1\alpha}^{\text{L}}\mathcal{C}_{1\beta}^{\text{R}}\hat{N}_1+
P^-Q^+\mathcal{C}_{2\alpha}^{\text{L}}\mathcal{C}_{2\beta}^{\text{R}}\hat{N}_2+
Q^-\mathcal{C}_{3\alpha}^{\text{L}}\mathcal{C}_{3\beta}^{\text{R}}\hat{N}_3
\big),
\qquad\text{(no summation over $\alpha$ and $\beta$)}
\label{transporter-SU3}
\end{equation}
where ($i=\text{L},\text{R}$)
\begin{gather}
M_1^i\ket{T_i^z,Y_i}=
\ket{T_i^z+\frac{1}{2},Y_i+\frac{1}{3}},\qquad
M_2^i\ket{T_i^z,Y_i}=
\ket{T_i^z-\frac{1}{2},Y_i+\frac{1}{3}},\qquad
M_3^i\ket{T_i^z,Y_i}=
\ket{T_i^z,Y_i-\frac{2}{3}},\\
P^{\pm}\ket{p,q}=\ket{p\pm1,q},\qquad Q^{\pm}\ket{p,q}=\ket{p,q\pm1},\qquad
T_i^{\pm}\ket{T_i}=\ket{T_i\pm\frac{1}{2}},\\
\mathcal{C}_{\alpha1}^i=T_i^+\hat{C}_{\alpha1}^{i(a)}+T_i^-\hat{C}_{\alpha1}^{i(b)},\qquad
\mathcal{C}_{\alpha2}^i=T_i^+\hat{C}_{\alpha2}^{i(a)}+T_i^-\hat{C}_{\alpha2}^{i(b)},\qquad
\mathcal{C}_{\alpha3}^i=\hat{C}_{\alpha3}^i,
\end{gather}
and the diagonal operators $\hat{N}_{1,2,3},\,
\hat{C}_{\alpha\beta}^{i(a,b)}$ and $\hat{C}_{\alpha3}^i$
can be found in~\cite{Kan-Nam_2107}.

As for the $SU(2)$ case, we can suppose that the $SU(3)$ theory has
mesons of the form of $\Phi_i$ in equations~\eqref{meson1-SU2}-\eqref{meson3-SU2},
but baryons are actually fermionic particles in this case,
\begin{equation}
\intp_4(x)=\frac{1}{a^{3/2}}\epsilon_{\alpha\beta\gamma}
\xi_{\alpha}^{\dagger}(x)\xi_{\beta}^{\dagger}(x)\xi_{\gamma}^{\dagger}(x).
\end{equation}

\section{Interpolating Operators on Qubits}
\label{4}
The theories and the operators discussed in the previous section
need to be mapped to a qubit system. In one dimension,
the Jordan-Wigner transformation works well to map
fermionic operators to qubit operators. More specifically
we choose the following orderings:
\begin{align}
\xi(x)\rightarrow\sigma^z(0)\cdots\sigma^z(x-1)\sigma^+(x)&
&&\text{for $U(1)$,}\\
\notag\\
\left.
\begin{aligned}
\xi_1(x)&\rightarrow\sigma_1^z(0)\sigma_2^z(0)\sigma_1^z(1)\cdots
\sigma_2^z(x-1)\sigma_1^+(x)\\
\xi_2(x)&\rightarrow\sigma_1^z(0)\sigma_2^z(0)\sigma_1^z(1)\cdots
\sigma_2^z(x-1)\sigma_1^z(x)\sigma_2^+(x)
\end{aligned}
\right\}&
&&\text{for $SU(2)$,}\\
\notag\\
\left.
\begin{aligned}
\xi_1(x)&\rightarrow\sigma_1^z(0)\sigma_2^z(0)\sigma_3^z(0)\sigma_1^z(1)\cdots
\sigma_3^z(x-1)\sigma_1^+(x)\\
\xi_2(x)&\rightarrow\sigma_1^z(0)\sigma_2^z(0)\sigma_3^z(0)\sigma_1^z(1)\cdots
\sigma_3^z(x-1)\sigma_1^z(x)\sigma_2^+(x)\\
\xi_3(x)&\rightarrow\sigma_1^z(0)\sigma_2^z(0)\sigma_3^z(0)\sigma_1^z(1)\cdots
\sigma_3^z(x-1)\sigma_1^z(x)\sigma_2^z(x)\sigma_3^+(x)
\end{aligned}
\right\}&
&&\text{for $SU(3)$,}
\end{align}
where $\sigma^{\pm}=(\sigma^x\pm i\sigma^y)/2$.

On the other hand, the local Hilbert space of the gauge field needs to be truncated.
For $U(1)$ this is quite simple:
if we use $k$ qubits for a single link, then we impose the cutoff
$\Lambda=2^{k-1}$ and we encode the states
$\ket{-\Lambda},\ket{-\Lambda+1},\dots,\ket{\Lambda-1}$
in the computational basis. The operator $U$ needs to be adapted
to the truncation. A common choice is to take
\begin{equation}
\begin{aligned}
&U\ket{\mathcal{E}}=
\ket{\mathcal{E}-1},\qquad\text{for }\mathcal{E}=-\Lambda+1,\dots,\Lambda-1,\\
&U\ket{-\Lambda}=\ket{\Lambda-1}.
\end{aligned}
\label{truncated-U}
\end{equation}
In this way $U$ is a permutation operator and can be
implemented on a qubit system using $O(k^3)$ gates and no ancillary qubit,
or using $O(k^2)$ gates and $k-1$ ancillary qubits.
For $SU(2)$ and $SU(3)$ the situation is more complicated but similar in essence.
We refer the reader to~\cite{Kan-Nam_2107} for the details.
After imposing a cutoff $\Lambda$, the permutation operators
$M_{\alpha}^{\text{L}/\text{R}},\,J^{\pm}$ of $SU(2)$, and
$M_{\alpha}^{\text{L}/\text{R}},\,P^{\pm},\,Q^{\pm},\,T_{\text{L}/\text{R}}^{\pm}$
of $SU(3)$, take the same form as $U$ in equation~\eqref{truncated-U}.

The diagonal operators $c_{\alpha\beta}^{\pm}$
in~\eqref{transporter-SU2} and the ones
in~\eqref{transporter-SU3}, like for instance
$\hat{C}_{11}^{\text{L}(a)}\hat{C}_{11}^{\text{R}(a)}\hat{N}_1$,
are not unitary. However, their entries lie between $-1$ and $1$, and
are efficiently computable classically. We can implement
them in the following way:
suppose that the function $f(x)$, where $x$ can take $M$ values
encoded in the computational basis, is
efficiently computable with classical techniques,
and take an ancillary register of $\eta=O(\log_2M)$ qubits
initialized in the state
\begin{equation}
\ket{F}=\frac{1}{\sqrt{N}}
\sum_{n=0}^{N-1}e^{-i2\pi n/N}\ket{n},\qquad N=2^{\eta}.
\end{equation}
Then we can build efficient quantum circuits to implement
unitary operators $U_f^{\pm}$ such that
\begin{equation}
U_f^{\pm}\ket{x}\ket{F}=
e^{\pm if(x)}\ket{x}\ket{F}.
\end{equation}
We can use LCU to combine the two operators and obtain
\begin{equation}
\frac{(U_f^+-U_f^-)}{2i}\ket{x}\ket{F}=
\sin[f(x)]\ket{x}\ket{F}.
\end{equation}
Going back to our case, denote by $\mathcal{D}$ a diagonal
operator, and by $\mathcal{D}(x)$ its entries, with
$x=(j,m_{\text{L}},m_{\text{R}})$
for $SU(2)$, and
$x=(p,q,T_{\text{L}},T_{\text{L}}^z,Y_{\text{L}},T_{\text{R}},T_{\text{R}}^z,Y_{\text{R}})$
for $SU(3)$. Then, taking $f(x)=\arcsin[\mathcal{D}(x)]$, we can implement $\mathcal{D}$.
In this way we can write $U_{\alpha\beta}$ as a linear combination
of unitary operators, and we can use LCU to implement it.

We can now write the interpolating operators in the qubit system,
together with the corresponding quantity $C$ defined
in equation~\eqref{LCU}.
For $U(1)$ they take the form
\begin{align}
\intp_1(x)&=\frac{1}{a}\big[\sigma^+(x)\sigma^-(x)-\sigma^+(x+1)\sigma^-(x+1)\big],
&\quad C_1&=\frac{2}{a},\\[3mm]
\intp_2(x)&=\frac{1}{a}\big[\sigma^-(x)U^{\dagger}(x)\sigma^+(x+1)+\sigma^-(x+1)U(x)\sigma^+(x)\big],
&\quad C_2&=\frac{2}{a},\\[3mm]
\intp_3(x)&=\frac{1}{a}\big[\sigma^-(x)U^{\dagger}(x)\sigma^+(x+1)-\sigma^-(x+1)U(x)\sigma^+(x)\big],
&\quad C_3&=\frac{2}{a}.
\end{align}

For $SU(2)$ they take the form
\begin{align}
\intp_1(x)&=\frac{1}{a}\sum_{\alpha}\big[\sigma_{\alpha}^+(x)\sigma_{\alpha}^-(x)-
\sigma_{\alpha}^+(x+1)\sigma_{\alpha}^-(x+1)\big],&\quad C_1&=\frac{4}{a},\\[1mm]
\intp_2(x)&=\frac{1}{a}\sum_{\alpha,\beta}\big[\sigma_{\alpha}^-(x)\zeta_{\alpha\beta}(x)U_{\alpha\beta}^{\dagger}(x)\sigma_{\beta}^+(x+1)+\sigma_{\alpha}^-(x+1)\zeta_{\alpha\beta}(x)U_{\alpha\beta}(x)\sigma_{\beta}^+(x)\big],&\quad C_2&=\frac{16}{a},\\[1mm]
\intp_3(x)&=\frac{1}{a}\sum_{\alpha,\beta}\big[\sigma_{\alpha}^-(x)\zeta_{\alpha\beta}(x)U_{\alpha\beta}^{\dagger}(x)\sigma_{\beta}^+(x+1)-\sigma_{\alpha}^-(x+1)\zeta_{\alpha\beta}(x)U_{\alpha\beta}(x)\sigma_{\beta}^+(x)\big],&\quad C_3&=\frac{16}{a},\\[1mm]
\intp_4(x)&=\frac{2}{a}\sigma_1^-(x)\sigma_2^-(x),&\quad C_4&=\frac{2}{a},
\end{align}
where
\begin{align}
\zeta_{11}(x)&=\sigma_2^z(x),&
\zeta_{12}(x)&=\sigma_2^z(x)\sigma_1^z(x+1),\\
\zeta_{21}(x)&=\mathbb{1},&
\zeta_{22}(x)&=\sigma_1^z(x+1),
\end{align}
and the $U_{\alpha\beta}$ are decomposed as in
equation~\eqref{transporter-SU2}.

For $SU(3)$ we have
\begin{align}
\intp_1(x)&=\frac{1}{a}\sum_{\alpha}\big[\sigma_{\alpha}^+(x)\sigma_{\alpha}^-(x)-
\sigma_{\alpha}^+(x+1)\sigma_{\alpha}^-(x+1)\big],&\quad C_1&=\frac{6}{a},\\[1mm]
\intp_2(x)&=\frac{1}{a}\sum_{\alpha,\beta}\big[\sigma_{\alpha}^-(x)\zeta_{\alpha\beta}(x)U_{\alpha\beta}^{\dagger}(x)\sigma_{\beta}^+(x+1)+\sigma_{\alpha}^-(x+1)\zeta_{\alpha\beta}(x)U_{\alpha\beta}(x)\sigma_{\beta}^+(x)\big],&\quad C_2&=\frac{130}{a},\\[1mm]
\intp_3(x)&=\frac{1}{a}\sum_{\alpha,\beta}\big[\sigma_{\alpha}^-(x)\zeta_{\alpha\beta}(x)U_{\alpha\beta}^{\dagger}(x)\sigma_{\beta}^+(x+1)-\sigma_{\alpha}^-(x+1)\zeta_{\alpha\beta}(x)U_{\alpha\beta}(x)\sigma_{\beta}^+(x)\big],&\quad C_3&=\frac{130}{a},\\[1mm]
\intp_4(x)&=\frac{6}{a^{3/2}}\sigma_1^z(0)\sigma_2^z(0)\sigma_3^z(0)\sigma_1^z(1)\cdots
\sigma_3^z(x-1)\sigma_1^-(x)\sigma_2^-(x)\sigma_3^-(x),&\quad C_4&=\frac{6}{a^{3/2}},
\end{align}
where again $\zeta_{\alpha\beta}(x)$ are strings of $Z$ Pauli
matrices analogous to the ones of the $SU(2)$ case,
and the $U_{\alpha\beta}$ are decomposed as in
equation~\eqref{transporter-SU3}. Notice that
the baryon is a fermionic particle, which implies that the
string of Pauli matrices $\sigma_1^z(0)\sigma_2^z(0)\sigma_3^z(0)\sigma_1^z(1)\cdots
\sigma_3^z(x-1)$ does not cancel out as for bosonic particles.

We have thus provided explicit linear expansions
of the interpolating operators in terms of
unitary operators on qubit systems. The number of
terms varies considerably from theory to theory and from
particle to particle, but in all cases it is a constant.
In the cases of $SU(2)$ and $SU(3)$, the quantities
$C_2$ and $C_3$ are quite large because of
the appearance of the operators $U_{\alpha\beta}$
in the expansion. This feature is specific of the Kogut-Susskind formulation.
The number of gates required for each term in the expansion
grows at most polynomially with the number of qubits dedicated
to a lattice site.

\section{Quantum Chromodynamics}
\label{5}
We focus now on two-flavoured quantum chromodynamics in three dimensions.
The first thing we need is to identify the continuum degrees of
freedom with the lattice degrees of freedom. This is what we do
in subsection~\ref{5A}. Then, in subsection~\ref{5B}
we discuss the wave-packet creation of pions and nucleons.

\subsection{Staggered fermions in three dimensions}
\label{5A}
\begin{figure}[ht]
\includegraphics{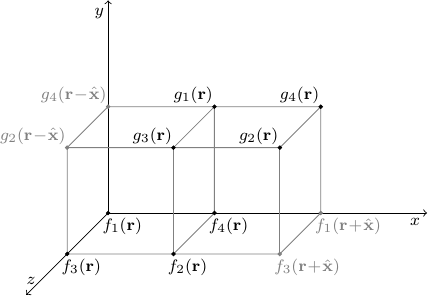}
\caption{\label{unit-cube1}Identification of the lattice
fields with the continuum fields.}
\end{figure}
The staggered-fermion Hamiltonian in three dimensions reads
\begin{equation}
H_0=\frac{1}{2a}\sum_{\mathbf{r},\hat{\mathbf{n}}}\big[
\xi^{\dagger}(\mathbf{r})\xi(\mathbf{r}+\hat{\mathbf{n}})+
\xi^{\dagger}(\mathbf{r}+\hat{\mathbf{n}})
\xi(\mathbf{r})\big]\eta(\mathbf{r},\hat{\mathbf{n}})+
m_0\sum_{\mathbf{r}}
(-1)^{x+y+z}\xi^{\dagger}(\mathbf{r})\xi(\mathbf{r})
\end{equation}
where $\mathbf{r}=(x,y,z)$ is a triplet of integers,
$\hat{\mathbf{n}}\in\{\hat{\mathbf{x}},\hat{\mathbf{y}},\hat{\mathbf{z}}\}$
is a unit vector along a positive Cartesian axis, and
\begin{equation}
\eta(\mathbf{r},\hat{\mathbf{x}})=(-1)^z,\qquad
\eta(\mathbf{r},\hat{\mathbf{y}})=(-1)^x,\qquad
\eta(\mathbf{r},\hat{\mathbf{z}})=(-1)^y.
\end{equation}
This Hamiltonian in the continuum limit describes two
flavours of quarks, up and down for example, with the same
mass $m_u=m_d=m_0$. Identifying the components of the two
Dirac spinors $u(\mathbf{r})$ and $d(\mathbf{r})$
with the field $\xi(\mathbf{r})$ is not trivial, and several
choices can be made up to $O(a)$ differences.
We show here two possibilities based
on~\cite{Susskind_1977-11,Banks-Raby-Susskind-Kogut-Jones-Scharbach-Sinclair_1977-02},
where they perform a similar identification in momentum space.

First modify $\xi(\mathbf{r})$ by a phase,
\begin{equation}
\phi(\mathbf{r})=i^{-x-z}A(y)D(x,z)\xi(\mathbf{r}),
\end{equation}
with
\begin{equation}
A(y)=\frac{1}{\sqrt{2}}\big[i^{y-1/2}+(-i)^{y-1/2}\big],\qquad
D(x,z)=\frac{1}{2}\big[(-1)^x+(-1)^z+(-1)^{x+z+1}+1\big].
\end{equation}
Then we identify $\phi$ at different sites
with  eight different fields as in figure~\ref{unit-cube1},
\begin{equation}
\begin{gathered}
f_1(\mathbf{r})=\phi(2\mathbf{r}),\quad
f_2(\mathbf{r})=\phi(2\mathbf{r}+\hat{\mathbf{x}}+\hat{\mathbf{z}}),\\
f_3(\mathbf{r})=\phi(2\mathbf{r}+\hat{\mathbf{z}}),\quad
f_4(\mathbf{r})=\phi(2\mathbf{r}+\hat{\mathbf{x}})\\
g_1(\mathbf{r})=\phi(2\mathbf{r}+\hat{\mathbf{x}}+\hat{\mathbf{y}}),\quad
g_2(\mathbf{r})=\phi(2\mathbf{r}+2\hat{\mathbf{x}}+\hat{\mathbf{y}}+\hat{\mathbf{z}}),\\
g_3(\mathbf{r})=\phi(2\mathbf{r}+\hat{\mathbf{x}}+\hat{\mathbf{y}}+\hat{\mathbf{z}}),\quad
g_4(\mathbf{r})=\phi(2\mathbf{r}+2\hat{\mathbf{x}}+\hat{\mathbf{y}}).
\end{gathered}
\label{phi-f-g-correspondence}
\end{equation}
The two Dirac spinors are obtained with the transformations
\begin{gather}
\tilde{u}=\frac{1}{\sqrt{2}a^{3/2}}(f+g),\qquad
\tilde{d}=\frac{M}{\sqrt{2}a^{3/2}}(f-g),\\
u=\frac{1}{\sqrt{2}}(\tilde{u}-i\tilde{d}),\qquad
d=\frac{1}{\sqrt{2}}(-i\tilde{u}+\tilde{d}),
\end{gather}
where
\begin{equation}
M=i
\begin{pmatrix}
\sigma_y & 0\\
0 & -\sigma_y
\end{pmatrix}
=
\begin{pmatrix}
0&1&0&0\\
-1&0&0&0\\
0&0&0&-1\\
0&0&1&0
\end{pmatrix}.
\end{equation}
We can put together the transformations as in
\begin{equation}
u=\frac{1}{2a^{3/2}}\big[(\mathbb{1}-iM)f+(\mathbb{1}+iM)g\big],\qquad
d=\frac{1}{2a^{3/2}}\big[(M-i\mathbb{1})f-(M+i\mathbb{1})g\big].
\label{u-d-identification}
\end{equation}

In terms of the fields $u(\mathbf{r})$ and $d(\mathbf{r})$,
the free fermion Hamiltonian reads
\begin{equation}
\begin{gathered}
H_0=\sum_{\mathbf{r}}a^3\bigg[
\frac{i}{2a}\sum_{\hat{\mathbf{n}}}\Big[
u^{\dagger}(\mathbf{r})\alpha_n\Delta_{\hat{\mathbf{n}}}u(\mathbf{r})+
d^{\dagger}(\mathbf{r})\alpha_n\Delta_{\hat{\mathbf{n}}}d(\mathbf{r})
\Big]
+\frac{i}{2a}\Big[
u^{\dagger}(\mathbf{r})\beta\gamma_5\Delta_{\hat{\mathbf{x}}\hat{\mathbf{y}}}
u(\mathbf{r})
-d^{\dagger}(\mathbf{r})\beta\gamma_5\Delta_{\hat{\mathbf{x}}\hat{\mathbf{y}}}
d(\mathbf{r})\Big]\\
+m_0\Big[
u^{\dagger}(\mathbf{r})\beta u(\mathbf{r})+
d^{\dagger}(\mathbf{r})\beta d(\mathbf{r})
\Big]\bigg]
\end{gathered}
\label{ud-Hamiltonian}
\end{equation}
where $\beta,\,\alpha_n$ and $\gamma_5$ are the usual
Dirac matrices in the Dirac representation,
\begin{equation}
\beta=
\begin{pmatrix}
\mathbb{1}	& 0\\
0			& -\mathbb{1}
\end{pmatrix},\quad
\alpha_n=
\begin{pmatrix}
0			& \sigma_n\\
\sigma_n	& 0
\end{pmatrix},\quad
\gamma_5=
\begin{pmatrix}
0			& \mathbb{1}\\
\mathbb{1}	& 0
\end{pmatrix},
\end{equation}
and
\begin{equation}
\begin{gathered}
\Delta_{\hat{\mathbf{x}}}q(\mathbf{r})=
\begin{pmatrix}
q_1(\mathbf{r}+\hat{\mathbf{x}})-q_1(\mathbf{r})\\
q_2(\mathbf{r})-q_2(\mathbf{r}-\hat{\mathbf{x}})\\
q_3(\mathbf{r}+\hat{\mathbf{x}})-q_3(\mathbf{r})\\
q_4(\mathbf{r})-q_4(\mathbf{r}-\hat{\mathbf{x}})
\end{pmatrix},
\quad
\Delta_{\hat{\mathbf{y}}}q(\mathbf{r})=\frac{1}{2}
\begin{pmatrix}
q_1(\mathbf{r})
-q_1(\mathbf{r}-\hat{\mathbf{y}})+
q_1(\mathbf{r}+\hat{\mathbf{x}}+\hat{\mathbf{y}})
-q_1(\mathbf{r}+\hat{\mathbf{x}})\\
q_2(\mathbf{r}+\hat{\mathbf{y}})
-q_2(\mathbf{r})+
q_2(\mathbf{r}-\hat{\mathbf{x}})
-q_2(\mathbf{r}-\hat{\mathbf{x}}-\hat{\mathbf{y}})\\
q_3(\mathbf{r})
-q_3(\mathbf{r}-\hat{\mathbf{y}})+
q_3(\mathbf{r}+\hat{\mathbf{x}}+\hat{\mathbf{y}})
-q_3(\mathbf{r}+\hat{\mathbf{x}})\\
q_4(\mathbf{r}-\hat{\mathbf{x}})
-q_4(\mathbf{r}-\hat{\mathbf{x}}-\hat{\mathbf{y}})+
q_4(\mathbf{r}+\hat{\mathbf{y}})
-q_4(\mathbf{r})
\end{pmatrix},\\
\Delta_{\hat{\mathbf{z}}}q(\mathbf{r})=
\begin{pmatrix}
q_1(\mathbf{r}+\hat{\mathbf{z}})-q_1(\mathbf{r})\\
q_2(\mathbf{r})-q_2(\mathbf{r}-\hat{\mathbf{z}})\\
q_3(\mathbf{r})-q_3(\mathbf{r}-\hat{\mathbf{z}})\\
q_4(\mathbf{r}+\hat{\mathbf{z}})-q_4(\mathbf{r})
\end{pmatrix},
\quad
\Delta_{\hat{\mathbf{x}}\hat{\mathbf{y}}}q(\mathbf{r})=
\frac{1}{2}
\begin{pmatrix}
q_1(\mathbf{r})
-q_1(\mathbf{r}-\hat{\mathbf{y}})
-q_1(\mathbf{r}+\hat{\mathbf{x}}+\hat{\mathbf{y}})
+q_1(\mathbf{r}+\hat{\mathbf{x}})\\
q_2(\mathbf{r}+\hat{\mathbf{y}})
-q_2(\mathbf{r})
-q_2(\mathbf{r}-\hat{\mathbf{x}})
+q_2(\mathbf{r}-\hat{\mathbf{x}}-\hat{\mathbf{y}})\\
q_3(\mathbf{r})
-q_3(\mathbf{r}-\hat{\mathbf{y}})
-q_3(\mathbf{r}+\hat{\mathbf{x}}+\hat{\mathbf{y}})
+q_3(\mathbf{r}+\hat{\mathbf{x}})\\
q_4(\mathbf{r}-\hat{\mathbf{x}})
-q_4(\mathbf{r}-\hat{\mathbf{x}}-\hat{\mathbf{y}})
-q_4(\mathbf{r}+\hat{\mathbf{y}})
+q_4(\mathbf{r})
\end{pmatrix}.
\end{gathered}
\label{discrete-derivatives}
\end{equation}
Assuming we can expand the fields in Taylor series,
in the continuum limit we have
\begin{equation}
\begin{gathered}
\Delta_{\hat{\mathbf{x}}}q(\mathbf{r})=
a\partial_xq(\mathbf{r})+O(a^2),\quad
\Delta_{\hat{\mathbf{y}}}q(\mathbf{r})=
a\partial_yq(\mathbf{r})+O(a^2),\quad
\Delta_{\hat{\mathbf{z}}}q(\mathbf{r})=
a\partial_zq(\mathbf{r})+O(a^2),\\
\Delta_{\hat{\mathbf{x}}\hat{\mathbf{y}}}q(\mathbf{r})=O(a^2),
\end{gathered}
\end{equation}
which confirms that the free staggered-fermion
Hamiltonian describes two species of Dirac fermions
in the continuum limit.

\begin{figure}[ht]
\includegraphics{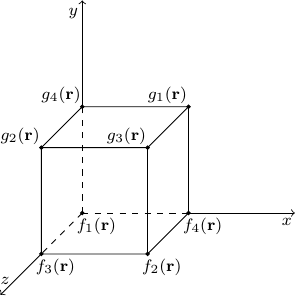}
\caption{\label{unit-cube2}Another identification of the lattice
fields with the continuum fields.}
\end{figure}

Another possible identification of the spinor components
is the following. Based on figure~\ref{unit-cube2},
we set
\begin{equation}
\begin{gathered}
f_1(\mathbf{r})=\phi(2\mathbf{r}),\quad
f_2(\mathbf{r})=\phi(2\mathbf{r}+\hat{\mathbf{x}}+\hat{\mathbf{z}}),\quad
f_3(\mathbf{r})=\phi(2\mathbf{r}+\hat{\mathbf{z}}),\quad
f_4(\mathbf{r})=\phi(2\mathbf{r}+\hat{\mathbf{x}})\\
g_1(\mathbf{r})=\phi(2\mathbf{r}+\hat{\mathbf{x}}+\hat{\mathbf{y}}),\quad
g_2(\mathbf{r})=\phi(2\mathbf{r}+\hat{\mathbf{y}}+\hat{\mathbf{z}}),\quad
g_3(\mathbf{r})=\phi(2\mathbf{r}+\hat{\mathbf{x}}+\hat{\mathbf{y}}+\hat{\mathbf{z}}),\quad
g_4(\mathbf{r})=\phi(2\mathbf{r}+\hat{\mathbf{y}}),
\end{gathered}
\end{equation}
and
\begin{equation}
u=\frac{1}{\sqrt{2}a^{3/2}}(f+g),\qquad
d=\frac{M}{\sqrt{2}a^{3/2}}(f-g).
\end{equation}
Then the Hamiltonian takes the form
\begin{equation}
\begin{gathered}
H_0=\sum_{\mathbf{r}}a^3\bigg[
\frac{i}{2a}\sum_{\hat{\mathbf{n}}}\Big[
u^{\dagger}(\mathbf{r})\alpha_n\Delta_{\hat{\mathbf{n}}}u(\mathbf{r})+
d^{\dagger}(\mathbf{r})\alpha_n\Delta_{\hat{\mathbf{n}}}d(\mathbf{r})
\Big]\\
+\frac{i}{2a}\Big[
u^{\dagger}(\mathbf{r})\beta\gamma_5\Delta_{\hat{\mathbf{x}}}^2
d(\mathbf{r})+
d^{\dagger}(\mathbf{r})\beta\gamma_5\Delta_{\hat{\mathbf{x}}}^2
u(\mathbf{r})+i
u^{\dagger}(\mathbf{r})\beta\gamma_5\Delta_{\hat{\mathbf{y}}}^2
d(\mathbf{r})-i
d^{\dagger}(\mathbf{r})\beta\gamma_5\Delta_{\hat{\mathbf{y}}}^2
u(\mathbf{r})\Big]\\
+m_0\Big[
u^{\dagger}(\mathbf{r})\beta u(\mathbf{r})+
d^{\dagger}(\mathbf{r})\beta d(\mathbf{r})
\Big]\bigg],
\end{gathered}
\end{equation}
where now
\begin{equation}
\begin{aligned}
\Delta_{\hat{\mathbf{x}}}q(\mathbf{r})&=
\frac{1}{2}
\big[q(\mathbf{r}+\hat{\mathbf{x}})-q(\mathbf{r}-\hat{\mathbf{x}})\big],
&\qquad
\Delta_{\hat{\mathbf{x}}}^2q(\mathbf{r})&=\frac{1}{2}
\big[q(\mathbf{r}+\hat{\mathbf{x}})+q(\mathbf{r}-\hat{\mathbf{x}})-
2q(\mathbf{r})\big],\\
\Delta_{\hat{\mathbf{y}}}q(\mathbf{r})&=\frac{1}{2}
\big[q(\mathbf{r}+\hat{\mathbf{y}})-q(\mathbf{r}-\hat{\mathbf{y}})\big],&\qquad
\Delta_{\hat{\mathbf{y}}}^2q(\mathbf{r})&=\frac{1}{2}
\big[q(\mathbf{r}+\hat{\mathbf{y}})+q(\mathbf{r}-\hat{\mathbf{y}})-
2q(\mathbf{r})\big],\\
\Delta_{\hat{\mathbf{z}}}q(\mathbf{r})&=
\begin{pmatrix}
q_1(\mathbf{r}+\hat{\mathbf{z}})-q_1(\mathbf{r})\\
q_2(\mathbf{r})-q_2(\mathbf{r}-\hat{\mathbf{z}})\\
q_3(\mathbf{r})-q_3(\mathbf{r}-\hat{\mathbf{z}})\\
q_4(\mathbf{r}+\hat{\mathbf{z}})-q_4(\mathbf{r})
\end{pmatrix}.
\end{aligned}
\end{equation}

The main difference between the two identifications
is that the first one is slightly more involved, but it
avoids terms that mix the two flavours in the free theory.
The second one is simpler, but the identified flavours are mixed.
We give them both because it is not clear which one is more convenient.
However, in the following, we consider only the first identification,
equations~\eqref{phi-f-g-correspondence}
and~\eqref{u-d-identification}, since it allows for a
better identification of the flavours.

\subsection{Interpolators in the Kogut-Susskind formulation}
\label{5B}
The Kogut-Susskind Hamiltonian for quantum chromodynamics on
a three-dimensional cubic lattice is
\begin{align}
H&=H_{\text{M}}+H_{\text{K}}+H_{\text{E}}+H_{\text{B}},\\
H_{\text{M}}&=m_0\sum_{\mathbf{r}}
(-1)^{x+y+z}\xi^{\dagger}(\mathbf{r})\xi(\mathbf{r}),\\
H_{\text{K}}&=\frac{1}{2a}\sum_{\mathbf{r},\hat{\mathbf{n}}}\big[
\xi^{\dagger}(\mathbf{r})U(\mathbf{r},\hat{\mathbf{n}})\xi(\mathbf{r}+\hat{\mathbf{n}})+
\xi^{\dagger}(\mathbf{r}+\hat{\mathbf{n}})U^{\dagger}(\mathbf{r},\hat{\mathbf{n}})
\xi(\mathbf{r})\big]\eta(\mathbf{r},\hat{\mathbf{n}}),\\
H_{\text{E}}&=\frac{g_0^2}{2a}\sum_{\mathbf{r},\hat{\mathbf{n}}}
E(\mathbf{r},\hat{\mathbf{n}})^2,\\
H_{\text{B}}&=-\frac{1}{2a^2g_0^2}\sum_{\square}\tr(P_{\square}+P_{\square}^{\dagger}),
\end{align}
where $P_{\square}$ is a plaquette operator.
The fields $\xi$, $U$ and $E$ are the $SU(3)$ fields
discussed in the previous section. The encoding of the
link Hilbert spaces does not differ from the one-dimensional case,
while for the fermions it is more convenient to use
a local encoding such as the generalized superfast
encoding~\cite{Setia-Bravyi-Mezzacapo-Whitfield_1810,Rhodes-Kreshchuk-Pathak_2405}. This encoding was given only for operators made of products
of an even number of fermionic operators. To create wave packets
of a fermionic particle, such as the nucleons, we need to
encode also odd fermionic operators. In the appendix we review
the generalized superfast encoding and extend it to odd operators.
In practice, the encoded versions of the operators
$\xi$ and $\xi^{\dagger}$ are made of operators $\sigma^{\pm}$
attached to a string of Pauli operators $\sigma^x$,
$\sigma^y$ and $\sigma^z$ on other qubits. This is qualitatively similar
to a Jordan-Wigner transformation,
except that the strings of Pauli operators have different shapes
depending on the details of the mapping.

Our goal here is to show how to write interpolating operators
for the particles appearing in QCD, like the pion and the nucleons,
in terms of qubit operators.
Knowing the correspondence between the fields $u$ and $d$
and $\xi$, equations~\eqref{phi-f-g-correspondence} and~\eqref{u-d-identification},
we can use results from the literature on
traditional lattice
calculations~\cite{Leinweber-Melnitchouk-Richards-Williams-Zanotti_0406,Basak-Edwards-Fiebig-Fleming-Heller-Morningstar-Richards-Sato-Wallace_0409,Burch-Gattringer-Glozman-Hagen-Lang-Schafer_0601,Burch-Gattringer-Glozman-Hagen-Hierl-Lang-Schafer_0604}.
However, the symmetries of the formulation used
in~\cite{Leinweber-Melnitchouk-Richards-Williams-Zanotti_0406,Basak-Edwards-Fiebig-Fleming-Heller-Morningstar-Richards-Sato-Wallace_0409,Burch-Gattringer-Glozman-Hagen-Lang-Schafer_0601,Burch-Gattringer-Glozman-Hagen-Hierl-Lang-Schafer_0604}
are different from the symmetries of the Kogut-Susskind
Hamiltonian. In particular, cubic and flavor symmetries
are intertwined between each other for staggered fermions, which
implies that spin and isospin mix together. On the other hand,
spin and isospin are among the main symmetries used to classify
and identify particles. As a consequence,
identification of particles in the Kogut-Susskind formulation
is possible only near the continuum limit, and up to $O(a)$
errors.

Pion interpolators are
\begin{equation}
\begin{gathered}
\intp_{\pi^+}(\mathbf{r})=d^{\dagger}(\mathbf{r})\gamma_0\gamma_5u(\mathbf{r}),
\qquad
\intp_{\pi^-}(\mathbf{r})=u^{\dagger}(\mathbf{r})\gamma_0\gamma_5d(\mathbf{r}),\\
\intp_{\pi^0}(\mathbf{r})=\frac{1}{\sqrt{2}}\big[
u^{\dagger}(\mathbf{r})\gamma_0\gamma_5u(\mathbf{r})
-d^{\dagger}(\mathbf{r})\gamma_0\gamma_5d(\mathbf{r})
\big].
\end{gathered}
\end{equation}
A proton interpolator is
\begin{equation}
\intp_{p}(\mathbf{r})=
\epsilon_{abc}\big[u_a^T(\mathbf{r})C\gamma_5d_b(\mathbf{r})\big]
u_c(\mathbf{r}),
\end{equation}
where $\mathcal{C}=-i\alpha_y$ is the charge-conjugation matrix
in the Dirac representation. The neutron interpolator
is obtained by exchanging $u$ and $d$.

These operators are not gauge invariant in the Kogut-Susskind
formulation, as they involve products between $\xi$ operators
at different lattice sites as a consequence of the identifications
in equation~\eqref{phi-f-g-correspondence}. We need to insert
appropriate link operators to make them gauge invariant.
The mesonic operators contain terms like
$\xi^{\dagger}(2\mathbf{r})\xi(2\mathbf{r}+\hat{\mathbf{x}})$,
which is made gauge invariant by
$\xi^{\dagger}(2\mathbf{r})
U(2\mathbf{r},\hat{\mathbf{x}})\xi(2\mathbf{r}+\hat{\mathbf{x}})$.
The baryonic operators contain terms like
$\epsilon_{abc}\xi_a(2\mathbf{r})\xi_b(2\mathbf{r}+\hat{\mathbf{x}})\xi_c(2\mathbf{r}+2\hat{\mathbf{x}}+\hat{\mathbf{y}})$,
which is made gauge invariant by
\begin{equation}
\epsilon_{abc}U_{aa'}(2\mathbf{r}+\hat{\mathbf{x}},-\hat{\mathbf{x}})\xi_{a'}(2\mathbf{r})
\xi_b(2\mathbf{r}+\hat{\mathbf{x}})
U_{cc'}(2\mathbf{r}+\hat{\mathbf{x}},\hat{\mathbf{y}})
U_{c'c''}(2\mathbf{r}+\hat{\mathbf{x}}+\hat{\mathbf{y}},\hat{\mathbf{x}})
\xi_{c'}(2\mathbf{r}+2\hat{\mathbf{x}}+\hat{\mathbf{y}}).
\end{equation}

The necessity of inserting link operators to restore
gauge invariance, together with the
transformations~\eqref{phi-f-g-correspondence}
and~\eqref{u-d-identification}, makes writing interpolators in the
Kogut-Susskind formulation extremely cumbersome.
This becomes particularly relevant when we calculate
the quantities $C$ for these operators:
We obtain approximately
$C_{\pi^{\pm}}\sim432/a^3$, $C_{\pi^0}=\sqrt{2}C_{\pi^{\pm}}$,
and $C_p\sim11500/a^{4.5}$. In a formulation such
as one with Wilson fermions, in which
each site contains the two spinors $u$ and $d$,
the corresponding quantities would be $C_{\pi^{\pm}}=12/a^3$,
$C_{\pi^0}=\sqrt{2}C_{\pi^{\pm}}$,
and $C_p=24/a^{4.5}$.

\section{Conclusions}
\label{6}
Quantum computation opens the possibility to study
real-time evolution of complex many-body quantum systems
in a general and systematic way. However, much work is required
on the experimental and theoretical sides to achieve
meaningful simulations that are inaccessible with traditional
techniques. In this manuscript, we aimed at solving
the
problem of preparing initial states suitable for
scattering simulation, consisting of two wave packets moving on top
of the vacuum. Simulating scattering of wave packets not only
allows for the computation of physical quantities to be
compared with experimental results, but also for disclosing
what happens inside the collision.

Our method is a practical application
of the Haag-Ruelle scattering theory developed
in the context of axiomatic quantum field theory more
than half a century ago.
In reference~\cite{Turco-Quinta-Seixas-Omar_2305}
we showed how to create wave packets from the vacuum
in scalar-field theories. Here we focused on lattice
gauge theories, and increased gradually the level of
difficulty, starting from a $U(1)$ theory in one dimension
and ending with QCD in three dimensions with two flavors of quarks.
We considered different kinds of particles, namely
pions and nucleons in QCD.

Three ingredients are necessary to apply our method. First,
we need to prepare the vacuum of the theory. This could be done
with variational methods,
like~\cite{Farrell-Illa-Ciavarella-Savage_2308,Davoudi-Hsieh-Kadam_2024-02},
or with an adiabatic
transformation~\cite{DAnna-Marinkovic-Barros_2024-11}.
Second, we need to know the spectrum of the theory. Near
the continuum limit, this amounts to know the particle content
and the particle masses. Finally, we need interpolating
operators for the particles we aim at. These operators are built
on a symmetry basis, and they should carry the quantum
numbers of the particles.
Our work is further extensible to other theories, formulations
and particles, with the only requirement that
there has to be a mass gap.

The creation of a wave packets
with our strategy requires
the use of LCU. This technique succeeds with
a probability $\rho$ that depends on the theory,
on the wave packet, and on the particle.
If we want to prepare $n$ wave packets,
the initial state preparations
needs to be repeated on average $O(1/\rho^n)$ times to obtain
the correct state for the simulation, so we can
create only a small number of incoming wave packets.
Fortunately, $n=2$ is the most relevant case.
The success probability seems to be the main limitation
of our approach, but at least it does not
affect the circuit depth, which is a favorable feature.

We chose to work in the Kogut-Susskind framework with staggered fermions.
For QCD, we imported interpolators for the pions and
the nucleons from the literature on traditional
lattice calculations. We transformed these operators
into operators in the staggered-fermion formulation,
but this raises some problems. There are many possible choices
for these transformations, with differences of the order
of the lattice spacing, and none of them respects
the symmetries that determine the particles
in the continuum. As a consequence, particle identification
is more problematic, a well-known downside of staggered fermions.
Furthermore,
the success probability suffers significantly from the
transformation of staggered fermions into Dirac spinors.
The interpolating operators in the staggered-fermion
formulation become long linear expansions of operators,
with insertion of gauge fields to achieve gauge invariance.
The resulting overheads in the success probability
are $\sim10^4$ for the pions, and $\sim10^8$ for the nucleons.
If we use Wilson fermions, the equivalent overheads are $\sim10^2$.

In the future,
the main issue that should be addressed is the success probability.
Finding a different method to implement the creation
operators with good probability would significantly
improve our approach. Furthermore,
it would be interesting to have detailed resource
estimates, especially in the
cases that have already been
tested~\cite{Farrell-Illa-Ciavarella-Savage_2401,Davoudi-Hsieh-Kadam_2024-02,Farrell-Zemlevskiy-Illa-Preskill_2025-05}
to make a comparison between the different strategies.
When the hardware is mature enough, it will be useful to test
the method on a real quantum computer.
It would also be important to have asymptotic estimates
of the other approaches in the literature to determine
which one is the most promising.

\section{Acknowledgments}
We are grateful for the support from FCT -- Funda\c{c}\~{a}o para a Ci\^{e}ncia e a Tecnologia (Portugal) -- namely through Project No. UIDB/04540/2020 and Contract LA/P/0095/ 2020, as well as from projects QuantHEP
and HQCC supported by the EU QuantERA ERA-NET Cofund in Quantum Technologies and by FCT (QuantERA/004/2021).
Finally, M. T. thanks FCT for support through
Grant No. PRT/BD/154668/2022. G. Q. acknowledges
the support from FCT, through projects CEECIND/
02474/2018, 2024.04456.CERN and FCT-Mobility/
1312232346/2024-25.

\appendix* \section{generalized superfast encoding of fermions}
\label{A}
We review the generalized superfast encoding (GSE) of fermions
into qubits~\cite{Setia-Bravyi-Mezzacapo-Whitfield_1810}.
In~\cite{Rhodes-Kreshchuk-Pathak_2405},
this encoding was used to estimate the resource requirements for
quantum simulation of lattice gauge theories such as quantum chromodynamics,
showing a substantial improvement with respect to a Jordan-Wigner transformation.
As presented in~\cite{Setia-Bravyi-Mezzacapo-Whitfield_1810},
the GSE allows for the representation of even fermionic operators only.
As odd operators are necessary for the preparation of fermionic-particle
wave packets, here we also show how they can be represented in the GSE.
In this appendix we drop the Einstein convention for repeated indices.

\subsection{Fermions}
We first introduce some general facts about fermions.
For a more detailed discussion see also~\cite{Bravyi-Kitaev_0003}.
Consider a system of $N$ independent fermionic modes with
annihilation operators $a_1,\dots,a_N$ satisfying the usual
anticommutation relations
\begin{equation}
\{a_I,a_J\}=0,\qquad\{a_I,a_J^{\dagger}\}=\delta_{IJ}.
\end{equation}
It is useful to introduce the Majorana modes as
\begin{equation}
\gamma_I^+=a_I+a_I^{\dagger},\qquad
\gamma_I^-=-i(a_I-a_I^{\dagger}).
\label{Majorana-modes-1}
\end{equation}
The Majorana modes are Hermitian, ${\gamma_I^{\pm}}^{\dagger}=\gamma_I^{\pm}$,
and satisfy the anticommutation relations
\begin{equation}
\{\gamma_I^{\rho},\gamma_J^{\sigma}\}=2\delta_{IJ}\delta_{\rho\sigma},
\label{Majorana-modes-2}
\end{equation}
where $\rho,\sigma\in\{+,-\}$. In particular, note that
${\gamma_I^{\pm}}^2=1$.

The Hilbert space of a fermionic system is spanned
by all possible polynomials in the
fermionic operators applied to the vacuum, and
has dimension $2^N$.
The vacuum is defined as
the state annihilated by all the annihilation operators,
\begin{equation}
a_I\ket{0}=0,\quad\forall I.
\end{equation}
There are two subspaces of dimension $2^{N-1}$ each, usually
called the even and the odd subspace. The even subspace
is spanned by arbitrary combinations of
products with an even number of fermionic operators
applied to the vacuum. Similarly, the odd subspace
is spanned by odd products of operators.
Combinations of states in different subspaces
are physically forbidden.

The even operators form a subalgebra generated by
\begin{equation}
B_I=-i\gamma_I^+\gamma_I^-,\qquad A_{IJ}=-i\gamma_I^+\gamma_J^+.
\label{even-generators}
\end{equation}
These operators satisfy the following relations:
\begin{gather}
B_I^{\dagger}=B_I,\qquad A_{IJ}^{\dagger}=A_{IJ},
\qquad B_I^2=1,\qquad A_{IJ}^2=1,\qquad A_{IJ}=-A_{JI},\label{even-properties-1}\\
B_IB_J=B_JB_I,\qquad
A_{IJ}B_K=(-1)^{\delta_{IK}+\delta_{JK}}B_KA_{IJ},\qquad
A_{IJ}A_{KL}=(-1)^{\delta_{IK}+\delta_{IL}+\delta_{JK}+\delta_{JL}}A_{KL}A_{IJ},
\label{even-properties-2}\\
A(\zeta)=
i^sA_{\zeta(1)\zeta(2)}A_{\zeta(2)\zeta(3)}\dots
A_{\zeta(s)\zeta(s+1)}=1,
\label{even-properties-3}
\end{gather}
where $\zeta:\,\{1,\dots,s+1\}\rightarrow\{1,\dots,N\}$ is a closed loop
of length $s$ with $\zeta(s+1)=\zeta(1)$.
Furthermore we have
\begin{align}
\prod_{I=1}^{N}B_I&=1& &\text{on the even subspace,}\\
\prod_{I=1}^{N}B_I&=-1& &\text{on the odd subspace.}
\end{align}

Given an arbitrary $\gamma_S^+$, we can write any other
Majorana mode as a product of $\gamma_S^+$ and an even operator:
\begin{equation}
\gamma_S^-=i\gamma_S^+B_S,\qquad
\gamma_I^+=i\gamma_S^+A_{SI},\qquad
\gamma_I^-=-\gamma_S^+A_{SI}B_I.
\end{equation}
This feature makes it very simple to implement
an arbitrary fermionic operator on a qubit system once
we have an encoding for the even operators.

\subsection{Even operators in the GSE}
To describe the GSE, we put the $N$ fermionic modes
on a connected graph with $N$ vertices. We also require that each
vertex be of even degree, and we assign an
orientation to each edge. The graph can be chosen freely to suite
a specific problem we want to address. For example, we might consider
a Hamiltonian $H$ and a graph $G=(V,E)$ such that
$H$ consists of terms acting nontrivially only on pairs
of nearest neighbours on $G$, namely
\begin{equation}
H=\sum_{(I,J)\in E}H_{IJ}.
\end{equation}

We denote by $d(I)$ the degree of vertex $I\in V$, and we
put $d(I)/2$ qubits on $I$. Next, we introduce
$d(I)$ local Majorana modes for each vertex,
$\gamma_{I1},\dots,\gamma_{Id(I)}$,
which satisfy the relations
\begin{align}
\{\gamma_{Ii},\gamma_{Ij}\}&=2\delta{ij},\\
[\gamma_{Ii},\gamma_{Jj}]&=0,\qquad\text{if $I\ne J$}.
\end{align}
We order the edges connected to vertex $I$ as
$1,\dots,d(I)$, and we attach the mode $\gamma_{Ii}$ to the edge $i$.
The local Majorana modes can be constructed with appropriate
strings of Pauli matrices acting on the qubits of vertex $I$.
From these modes we define the operators
\begin{align}
\tilde{B}_I&=(-i)^{d(I)/2}\gamma_{I1}\cdots\gamma_{Id(I)}&
\qquad&\text{for each vertex $I$,}\label{def-B-tilde}\\
\tilde{A}_{IJ}&=\epsilon_{IJ}\gamma_{Ii}\gamma_{Jj}&
\qquad&\text{for each edge $(I,J)$,}
\label{def-A-tilde}
\end{align}
where $\epsilon_{IJ}=1$ if we follow the arrow when going from
$I$ to $J$, and $\epsilon_{IJ}=-1$ if we go in the opposite direction.
The modes $\gamma_{Ii}$ and $\gamma_{Jj}$
in~\eqref{def-A-tilde} are the two modes attached to the
ends of edge $(I,J)$.

The operators defined in the expressions~\eqref{def-B-tilde} and
\eqref{def-A-tilde} satisfy
the properties~\eqref{even-properties-1} and \eqref{even-properties-2},
but not the property~\eqref{even-properties-3}, so we
need to impose it.
We define the qubit counterparts of the loop operators
in~\eqref{even-properties-3} as
\begin{equation}
\tilde{A}(\zeta)=
i^s\tilde{A}_{\zeta(1)\zeta(2)}\tilde{A}_{\zeta(2)\zeta(3)}\dots
\tilde{A}_{\zeta(s)\zeta(s+1)}
\end{equation}
for any closed loop $\zeta$ of length $s$ on the graph $G$,
where now $\zeta(i)$ and $\zeta(i+1)$ are connected by an edge.
The group $\mathcal{S}$ generated by these loop operators
form an Abelian group that commutes with
$\tilde{A}_{IJ}$ and $\tilde{B}_I$, and can be used
as a stabilizer of the subspace
\begin{equation}
\mathcal{L}=\{\ket{\psi}:\tilde{A}(\zeta)\ket{\psi}=\ket{\psi}
\text{ for all loops }\zeta\}.
\end{equation}
We can identify either the even or the odd subspace
of the fermionic Hilbert space with the stabilized space $\mathcal{L}$,
which has dimension $2^{N-1}$. The operators
$\tilde{A}_{IJ}$ and $\tilde{B}_I$, when restricted to this subspace,
satisfy also the property~\eqref{even-properties-3}
by construction of $\mathcal{L}$, and give us a qubit representation of
the algebra of even operators.
Since $G$ is a connected even-degree graph, there exists an
Eulerian cycle, namely a loop $\eta$ that utilizes every edge
of the graph exactly once. For this loop we have
\begin{equation}
\prod_{I=1}^N\tilde{B}_I=\pm\tilde{A}(\eta),
\label{parity-representation}
\end{equation}
where the sign depends on our choice of the orientation of the
edges. If our orientation produces a positive sign,
$\mathcal{L}$ represents the even subspace. Then,
if we switch the direction of one arrow, we obtain
a different subspace $\mathcal{L}$, which represents the odd subspace.

\subsection{Odd operators in the GSE}
In this section we show how to extend the generalized superfast
encoding described in the previous section to also represent
odd fermionic operators. Given a graph, a specific orientation
of the edges gives one of many possible stabilized subspaces.
Each of them can be identified either with the
even or the odd subspace of the fermionic Hilbert space.
Even operators, and their qubit representations,
preserve these subspaces, while odd
operators do not in the fermionic space and neither
should their representation on the qubit space.
Our goal is to find qubit operators representing
the Majorana modes in~\eqref{Majorana-modes-1} and~\eqref{Majorana-modes-2}.
This representation should be compatible with
the representation $\tilde{A}_{IJ}$ and $\tilde{B}_{IJ}$
of the previous section, at least on the stabilized subspace,
and should make us jump from a space where
equation~\eqref{parity-representation} holds
with the positive sign, to a space where it holds
with the negative sign, and vice versa.

Consider again a graph as in the previous section, with every vertex
of even degree, and take two vertices $S$ and $T$ connected by an edge.
For every vertex $I$ in the graph take a path $\zeta_{SI}$
from $S$ to $I$ that does not use the edge $(S,T)$,
with $\zeta_{SI}(1)=S$ and $\zeta_{SI}(s)=I$. We can always
find such paths following the Eulerian cycle $\eta$.
Consider now the local Majorana mode on $S$
attached to the edge $(S,T)$, which we can call $\gamma_{S1}$
(up to relabeling of the modes). Motivated by
\begin{align}
\gamma_I^+&=i\gamma_S^+A_{SI},\\
\gamma_I^-&=-\gamma_S^+A_{SI}B_I,
\end{align}
as a representation of the fermionic operators we take
\begin{align}
\gamma_I^+&\rightarrow\hat{\gamma}_I^+=\gamma_{S1}\tilde{A}(\zeta_{SI}),\\
\gamma_I^-&\rightarrow\hat{\gamma}_I^-=i\gamma_{S1}\tilde{A}(\zeta_{SI})\tilde{B}_I,
\end{align}
where
\begin{align}
\tilde{A}(\zeta_{SS})&=1,\\
\tilde{A}(\zeta_{SI})&=i^{s-1}\tilde{A}_{\zeta_{SI}(1)\zeta_{SI}(2)}
\tilde{A}_{\zeta_{SI}(2)\zeta_{SI}(3)}\dots\tilde{A}_{\zeta_{SI}(s-1)\zeta_{SI}(s)}.
\end{align}
To see that this is a good representation of $\gamma_I^{\pm}$, first notice that
\begin{align}
&[\gamma_{S1},\tilde{A}_{ST}]=0,\label{SJ-commutator}\\
&\{\gamma_{S1},\tilde{A}_{SJ}\}=0\qquad\text{if $J\ne T$}.
\label{SJ-anticommutator}
\end{align}
The difference between the first and the second line is
the reason for our choice of the paths $\zeta_{SI}$.
Then, from~\eqref{even-properties-1}, \eqref{even-properties-2}
and~\eqref{SJ-anticommutator}, it is easy to see
that $\hat{\gamma}_I^{\pm}$ satisfy the same anticommutation relations
as $\gamma_I^{\pm}$.

The generators of the even algebra obtained from $\hat{\gamma}_I^+$
and $\hat{\gamma}_I^-$ are
\begin{equation}
\hat{B}_I=-i\hat{\gamma}_I^+\hat{\gamma}_I^-,\qquad
\hat{A}_{IJ}=-i\hat{\gamma}_I^+\hat{\gamma}_J^+,
\end{equation}
and it is straightforward to see that
$\hat{B}_I=\tilde{B}_I$.
Suppose now there is an edge $(I,J)$. In general, we have
\begin{gather}
\hat{A}_{IJ}\ne\tilde{A}_{IJ}.
\end{gather}
However, on the subspace
$\mathcal{L}=\{\ket{\psi}:\tilde{A}(\zeta)\ket{\psi}=\ket{\psi}\}$, we have
\begin{equation}
\hat{A}_{IJ}\ket{\psi}=\hat{A}_{IJ}\tilde{A}(\zeta)\ket{\psi}.
\end{equation}
If we choose the loop $\zeta$ as the concatenation of $\zeta_{SI}$, $(I,J)$
and the reverse of $\zeta_{SJ}$,
we obtain
$\hat{A}_{IJ}\tilde{A}(\zeta)=\tilde{A}_{IJ}$,
and we can conclude
\begin{equation}
\hat{A}_{IJ}\ket{\psi}=\tilde{A}_{IJ}\ket{\psi}\quad
\forall\ket{\psi}\in\mathcal{L}.
\end{equation}
From this we can see that the representation of the
even algebra $\tilde{B}_I$ and $\tilde{A}_{IJ}$
is embedded in many representations of
$\hat{\gamma}_I^+$ and $\hat{\gamma}_I^-$,
one for each possible choice of the paths $\zeta_{SI}$.
When we restrict these representations to the stabilized space
$\mathcal{L}$, they all coincide and give the representation
of the even algebra introduced in~\cite{Setia-Bravyi-Mezzacapo-Whitfield_1810}.

To conclude, we need to see what happens when we act
with an odd operator on a state in the stabilized subspace $\mathcal{L}$.
Since all the operators $\tilde{A}_{IJ}$ and $\tilde{B}_I$
commute with the stabilizer operators $\tilde{A}(\zeta)$, the
relations~\eqref{SJ-commutator} and~\eqref{SJ-anticommutator}
imply
\begin{align}
&\{\hat{\gamma}_I^{\pm},\tilde{A}(\zeta)\}=0&&
\text{if the loop $\zeta$ involves $(S,T)$;}
\label{loop-relations1}\\
&[\hat{\gamma}_I^{\pm},\tilde{A}(\zeta)]=0&&
\text{if the loop $\zeta$ does not involve $(S,T)$.}
\label{loop-relations2}
\end{align}
Suppose now that our initial choice of the arrows
in the graph allows us to identify the space stabilized
by $\{\tilde{A}(\zeta)\}$ with the even subspace,
\begin{equation}
\mathcal{L}^+=\{
\ket{\psi}:\tilde{A}(\zeta)\ket{\psi}=\ket{\psi}
\text{ for all loops }\zeta\},\qquad
\prod_{I=1}^N\tilde{B}_I=\tilde{A}(\eta)
\end{equation}
for an Eulerian cycle $\eta$.
If we define
\begin{align}
\bar{A}_{ST}&=-\tilde{A}_{ST},\\
\bar{A}_{IJ}&=\tilde{A}_{IJ}\qquad\text{if $\{I,J\}\ne\{S,T\}$},
\end{align}
-- which is equivalent to switching the arrow of $(S,T)$ --
then $\{\bar{A}(\zeta)\}$ is a stabilizer for the odd subspace,
\begin{equation}
\mathcal{L}^-=\{
\ket{\psi}:\bar{A}(\zeta)\ket{\psi}=\ket{\psi}
\text{ for all loops }\zeta\},\qquad
\prod_{I=1}^N\tilde{B}_I=-\bar{A}(\eta).
\end{equation}
As a consequence of the relations~\eqref{loop-relations1}
and~\eqref{loop-relations2}, any odd operator
makes us jump from $\mathcal{L}^+$ to $\mathcal{L}^-$, and vice versa.
In virtue of this, we can conclude that the operators
$\hat{\gamma}_I^{\pm}$, together with the two
spaces $\mathcal{L}^{\pm}$ provide us with a full representation
of the fermionic Hilbert space and the associated algebra.

\subsection{GSE for hypercubic lattices}
\begin{figure}[hb]
\includegraphics{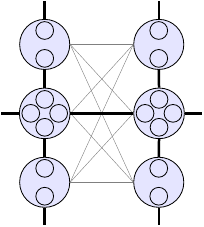}
\caption{\label{example-SU3}GSE graph for $SU(3)$ on a square lattice.
The gray lines represent physical interaction terms in
the Hamiltonian, and there is no such line between
modes on the same site. The thick lines are edges
of the graph we use for our encoding, and we also show
the associated local Majorana modes on each mode.}
\end{figure}

In this work we use the GSE for $SU(N)$ gauge theories on
hypercubic lattices with one flavour of fermions, which means that we have
$N$ independent fermionic modes on each site.
In the one-dimensional case, the GSE essentially reduces to the
Jordan-Wigner transformation. In $d$ dimensions we use $N+d-1$ qubits per lattice
site. As an example, figure~\ref{example-SU3} depicts two neighbouring sites
in an $SU(3)$ theory on a square lattice.
The gray lines represent physical interaction terms in
the Hamiltonian, and there is no such line between
modes on the same site. The thick lines are edges
of the graph we use for our encoding, and we also show
the associated local Majorana modes on each mode.
To obtain an interaction term between two modes, we simply
choose the shortest path on the graph connecting the two modes,
and the resulting encoded operator is in general
a string of operators acting on the qubits involved in the path.
Two out of three fermionic modes use one qubit each, the other modes
use two qubits each. If mode $I$ uses one qubit,
we take as local Majorana modes
\begin{equation}
\gamma_{I1}=\sigma^x,\qquad\gamma_{I2}=\sigma^y.
\end{equation}
If mode $I$ uses two qubits, we take
\begin{equation}
\gamma_{I1}=\sigma^x\otimes\mathbb{1},\qquad\gamma_{I2}=\sigma^y\otimes\mathbb{1},\qquad
\gamma_{I3}=\sigma^z\otimes\sigma^x,\qquad\gamma_{I4}=\sigma^z\otimes\sigma^y.
\end{equation}
Adaptation of this example to an $SU(3)$ theory on a cubic lattice
is straightforward.

\bibliography{Bibliography_Matteo-Turco}

\end{document}